\documentclass[journal=jpc,manuscript=article]{achemso}
\usepackage[utf8]{inputenc}

\usepackage{longtable}
\usepackage{multirow}
\usepackage{amsmath}
\usepackage{subfigure}
\usepackage[usenames,dvipsnames]{xcolor}
\usepackage{soul}
\usepackage{bm}

\usepackage{xr}
\usepackage{amssymb}
\usepackage{siunitx}
\usepackage{multirow}
\usepackage{multicol} \usepackage{wrapfig} \usepackage{enumitem}
\usepackage{bm} \usepackage{outlines} 
\usepackage{url}

\usepackage{siunitx}\usepackage{booktabs} 
\usepackage{soul}

\newcommand{\mcc}[0]{{\bf M-CC}}
\newcommand{\mdft}[0]{{\bf M-DFT}}
\newcommand{\mccs}[0]{{\bf M-CC }}
\newcommand{\mdfts}[0]{{\bf M-DFT }}

\usepackage[normalem]{ulem}

\makeatletter
\newcommand*{\addFileDependency}[1]{
  \typeout{(#1)}
  \@addtofilelist{#1}
  \IfFileExists{#1}{}{\typeout{No file #1.}}
}
\makeatother

\newcommand*{\myexternaldocument}[1]{%
    \externaldocument{#1}%
    \addFileDependency{#1.tex}%
    \addFileDependency{#1.aux}%
}

\myexternaldocument{si}
\SectionNumbersOn

\author{Eric D. Boittier} \affiliation[University of Basel]{Department
  of Chemistry, University of Basel, Klingelbergstrasse 80, CH-4056
  Basel, Switzerland.}
  
\author{Silvan K\"aser} \affiliation[University of Basel]{Department
  of Chemistry, University of Basel, Klingelbergstrasse 80, CH-4056
  Basel, Switzerland.}
  
  \author{Markus Meuwly} \affiliation[University of Basel]{Department
    of Chemistry, University of Basel, Klingelbergstrasse 80, CH-4056
    Basel, Switzerland.}  \email{m.meuwly@unibas.ch}

\title{Towards Large-Scale Condensed Phase Simulations using Machine
  Learned Energy Functions}

\begin{document}
\date{\today}

\begin{abstract}
Accurate, yet computationally efficient energy functions are essential
for state-of-the art molecular dynamics (MD) studies of condensed
phase systems. Here, a generic workflow based on a combination of
machine learning-based and empirical representations of intra- and
intermolecular interactions is presented. The total energy is
decomposed into internal contributions, and electrostatic and van der
Waals interactions between monomers. The monomer potential energy
surface is described using a neural network, whereas for the
electrostatics the flexible minimally distributed charge model is
employed. Remaining contributions between reference energies from
electronic structure calculations and the model are fitted to standard
Lennard-Jones (12-6) terms. For water as a topical example, reference
energies for the monomers are determined from CCSD(T)-F12 calculations
whereas for an ensemble of cluster structures containing $[2,60]$ and
$[2,4]$ monomers DFT and CCSD(T) energies, respectively, were used to
best match the van der Waals contributions. Based on the bulk liquid
density and heat of vaporization, the best-performing set of LJ(12-6)
parameters was selected and a wide range of condensed phase properties
were determined and compared with experiment. MD Simulations on the
multiple-nanosecond time scale were carried out for water boxes
containing 2000 to 8000 monomers, depending on the property
considered. The performance of such a generic ML-inspired
parametrization scheme is very promising and future improvements and
extensions are discussed, also in view of recent advances for water in
particular in the literature.
\end{abstract}

\section{Introduction}
Molecular dynamics (MD) simulations have long been a cornerstone in
fields ranging from chemistry and biology to materials science and
drug
discovery.\cite{van1990computer,karplus1990molecular,durrant2011molecular}
These simulations offer a unique lens through which to explore the
dynamic behaviour of molecules and materials at the atomic scale and
to shed light on molecular interactions and complex physical and
chemical processes. Central to the success of MD simulations is the
choice of model to describe the total energy of a molecular system at
the atomic level that governs the dynamics of the
system. Traditionally, force fields such as the CHARMM general force
field (CGenFF)\cite{cgenff:2010}, AMBER,\cite{amberff}
OPLS,\cite{oplsff} or GROMOS\cite{gromos:2004} energy functions have
relied largely on empirical mathematical functions whereby parameters
were fitted to experimental data wherever possible. While traditional
empirical force fields have played a crucial role in advancing the
understanding of molecular systems, they are not without
limitations. One of the most notable challenges lies in balancing
accuracy and computational efficiency. This trade-off between accuracy
and speed has been a persistent bottleneck in the field of molecular
dynamics.\\

\noindent
With the advent of machine learning (ML) based techniques, new
possibilities for conceiving potential energy surfaces for individual
molecules and simulations in the condensed phase emerge. Typical
empirical energy functions (EEFs) use harmonic springs to represent
bonds and valence angles, and periodic functions for
dihedrals. However, for high-quality simulations of vibrational
spectroscopy or energy transfer it is necessary to go beyond the
harmonic approximation. Effects such as mechanical anharmonicity and
coupling between different internal degrees of freedom can be included
through the use of ML
approaches.\cite{MM.n3:2019,MM.jcp:2020,MM.rkhs:2020,nandi:2019,li:2014}
For the nonbonded interactions it is common within standard EEFs to
employ atom-centered point charges and a Lennard-Jones (LJ)
representation for van-der-Waals (vdW)
interactions.\cite{mackerell:2004} Again, to represent chemically
relevant features such as sigma-holes or lone pairs, it is imperative
to modify the functional form, for example through the use of
atom-centered multipoles.\cite{Stone2013} Including higher-order
atomic multipoles improves the accuracy but at the expense of
increased computational cost and implementation
complexity.\cite{Handley2009,MM.mtp:2013,Devereux2014,bereau:2016}
Additional important contributions arise from
polarization\cite{ren:2019} and charge-transfer
contributions\cite{sidler:2018} all of which can influence the
energetics, dynamics, and spectroscopy in the condensed
phase. Finally, from an EEF-perspective the vdW interactions are often
represented as LJ terms with {\it ad hoc} (Lorentz-Berthelot)
combination rules. Alternative and potentially improved
representations are the buffered 14-7
parametrization\cite{halgren:1992} and/or modified combination
rules.\cite{mason:1988,millie:2001}\\

\noindent
Capitalizing on the toolbox of ML-based representations for energies
and forces, the question arises how to organize a robust and
extensible workflow for an initial model suitable for condensed-phase
simulations.\cite{Lemkul:2025,Mauger:2025} Firstly, such a workflow
should describe the ``bonded terms'' (or internal degrees of freedom)
for the monomers of the system considered at the highest achievable
level of quantum chemical theory while remaining practical and
computationally efficient in its
evaluation.\cite{Heindel:2025,Teng:2025} Possibilities are kernel- or
neural network- (NN) based representations, or a combination of the
two.\cite{MM.rkhs:2017,MM.physnet:2019,MM.kernn:2024} With regards to
the level of theory, coupled-cluster quality can be achieved through
transfer learning (TL) which provides an attractive route for
high-level energy functions for medium-sized
molecules.\cite{MM.tl:2021,MM.tl:2022,MM.tl:2025} Secondly, for the
nonbonded interactions representing the electrostatic potential
around each monomer is essential, in particular if monomers carry
chemically relevant features such as sigma-holes or lone pairs which
can also depend on monomer geometry. ML-based approaches for encoding
such information have been recently developed and include
multipoles,\cite{bereau:2015} kernel-based minimally distributed
charges (kMDCM),\cite{MM.kmdcm:2024} or fluctuating
charges.\cite{ko:2021}\\

\noindent
As with all ML-based tasks, choosing a suitable training set is as
relevant as the concrete model chosen. For applications in chemistry a
balance between physical realism and computational feasibility needs
to be struck. One possibility is to use clusters of monomers which has
already been done to conceive a fluctuating charge model for
water.\cite{banks:1999} To parametrize vdW contributions, clusters of
different sizes extracted from MD simulations and using their total
interaction energies for fitting was done for a range of
systems.\cite{MM.ff:2024} This is also a meaningful approach for
electrostatically demanding systems, such as eutectic
liquids.\cite{MM.eutectic:2024, MM.eutectic:2025} Using reference data
from density functional theory calculations, specific force field
parameters were adjusted to best reproduce the reference data. One
attractive aspect for such an approach is that, {\it a priori}, no
experimentally measured data is required. On the other hand,
comparison with measured data provides a route for further improving
the quality of the energy function.\\

\noindent
Water as a paradigmatic ``complex liquid'' is an ideal system for
developing a ML-based workflow. Accurately capturing water’s unique
gas and solution phase properties is a formidable challenge making the
development of accurate and still computationally efficient water
models key for advancing research, e.g. in biomolecular and materials
sciences. Research in water modeling for MD simulations generally
follows two main directions. The first aims to capture water
interactions with high precision by incorporating many-body
interactions (up to four-body).\cite{yu:2022,zhu2023mb} These models
offer an exceptionally accurate description of water but are often
limited in their application to smaller system sizes (typically 256
monomers) due to their high computational cost. On the other hand,
more scalable models were designed to handle large system sizes
efficiently by fitting the model parameters to best reproduce
experimentally determined quantities such as density, heat of
vaporization and self-diffusion.\cite{jorgensen:1983}\\

\noindent
The present work combines a CCSD(T)-F12 monomer PES for the internal
(intramolecular) degrees of freedom and an accurate description of the
intermolecular interactions based on advanced electrostatic models and
Lennard-Jones interactions fitted to interaction energies for clusters
of different sizes. Such a model is shown to achieve accurate
gas-phase spectroscopic properties, to qualitatively reproduce the
structure and energetics of small water clusters, and to yield good
bulk properties from MD simulations based on box sizes with $\sim
10^4$ water molecules compared with experiments. For the
intermolecular interactions the impact of the chosen level of theory
was assessed by training independent models on reference interaction
energies at the DFT and CCSD(T) levels. By avoiding computationally
demanding many-body terms, it remains efficient enough for large-scale
simulations. While this work focuses solely on water, for which many
experimental observables have been determined for comparison and
eventual refinement, the main thrust of the work is to establish a
rational route for developing accurate and efficient models for other
condensed phase systems for which the relevant experimental
observables are often unavailable.\\

\section{Methods}
The computational model considered here is built on two 
ML-based components. One of them represents the intramolecular
potential of a monomer and the other describes the intermolecular
interactions between monomers. The theoretical foundations for both
components and the data required to determine them are introduced
below, followed by details regarding the molecular dynamics (MD)
simulations and their analysis.

\subsection{Intramolecular Model}
Reference data was obtained for the water monomer by running $NVT$ MD
simulations at 300 and 1500~K using the semiempirical GFN2-xTB
method\cite{bannwarth2019gfn2} (2501 structures). These were augmented
with structures along an O--H bond stretch (15 structures) that covers
$1.0 < r_{\rm O-H} < 1.70$~\AA~and along the HOH bend covering 80 to
$130^\circ$ (50 structures). The reference energies and forces were
then calculated for a total of 2566 water configurations at the
CCSD(T)-F12B/aug-cc-pVTZ-F12 level of theory (\mcc) using
MOLPRO.\cite{MOLPRO,adler2007simple}\\

\noindent
The total internal energy of a single water molecule was represented
by a small and fully connected feed-forward NN. The basic building
blocks of the NN are dense layers
\begin{align}
    y = \bm{Wx} + \bm{b},
\end{align}
which need to be stacked and combined with a non-linear activation
function $\sigma$ to be able to model non-linear relationships. The
employed NN is represented mathematically in
Equation~\ref{eq:ffnet}. Here, the activation function $\sigma$ was a
soft plus function, and the last layer was a linear
transformation. The input to the NN are the three interatomic
distances and the output is the potential energy $E$ of the
molecule. The forces $\bm{F}_i = \frac{\partial E}{\partial \bm{r}_i}$
acting on the atoms are obtained from reverse mode automatic
differentiation.\cite{baydin2017automatic}
\begin{align}\label{eq:ffnet}
    V = \bm{W_3}\sigma(\bm{W_2}\sigma(\bm{W_1}\sigma(\bm{W_0r} +
    \bm{b_0}) + \bm{b_1}) + \bm{b_2}) + b_3
\end{align}
The parameters of the NN are optimized by minimizing an appropriate
loss function $\mathcal{L}$ using ADAM\cite{kingma2014adam}. The loss
function incorporates both the total energy and the forces and is
given by
\begin{align}
    \mathcal{L} = \left|E - E^{\rm ref}\right| + \omega_F \sum_{i=1}^N
    \sum_{\alpha=1}^3\left| - \frac{\partial E}{\partial r_{i,\alpha}}
    - F^{\rm ref}_{i,\alpha} \right|.
\end{align}
$E$ and $E^{\rm ref}$ correspond to the model and reference energies,
$F^{\rm ref}_{i,\alpha}$ are the Cartesian components of the reference
force on atom $i$, and $r_{i,\alpha}$ is the $\alpha$th Cartesian
coordinate of atom $i$. The hyperparameter $\omega_F$ weighs the
contribution of the forces to the total loss function and was set to
$\omega_F = 10$.\cite{MM.physnet:2019}\\

\noindent
Rotational and translational invariances are ensured by employing
internal coordinates (interatomic distances) but permutational
invariance of like atoms was not incorporated explicitly for the
benefit of computational efficiency. This, however, could easily be
achieved by, e.g. following an approach based on fundamental
invariants.\cite{shao2016communication} The simplicity of the NN used
in the present work enables the implementation of the NN in FORTRAN,
and thus in CHARMM. Ultimately this yields a significant speed-up as
well as accessibility to many functionalities that have been
implemented in CHARMM during the last decades.\cite{MM.charmm:2024}
The PES for the water monomer builds on recent work that promotes the
use of very simple models, with additional technical details provided
elsewhere.\cite{MM.kernn:2024} The primary difference here is that
interatomic distances are used directly as descriptors instead of
reproducing kernels because the energy function does not need to be
reactive.\cite{rabitz:1999,MM.rkhs:2017} In the context of the most
advanced energy functions for water suitable for (small scale)
condensed phase simulations - MB-Pol\cite{reddy:2016} and
q-AQUA\cite{yu:2022} - this contribution is described by the highly
accurate empirically adjusted Schwenke/Partridge 1-body
term.\cite{schwenke:1997}\\

\subsection{Intermolecular Model}
The parametrization strategy for the nonbonded (intermolecular)
contributions followed recent work and is based on interaction
energies of small- to medium sized clusters.\cite{MM.ff:2024} To
generate the sample structures, MD simulations using
CHARMM\cite{charmm:2009,MM.charmm:2024} for a cubic box of size 35
\AA\/ containing 1000 water molecules and using periodic boundary
conditions. The bonded (internal) energies were computed with an
existing RKHS-based representation\cite{fischer2023first,MM.ff:2024}
and the nonbonded interactions were those from the TIP3P
model\cite{jorgensen:1983} with fluctuating minimal distributed
charges (fMDCM)\cite{MM.fmdcm:2022}, although both choices are not
critical except for the fact that the model needs to be valid for
deformation of the water monomers.\\

\noindent
The simulations were carried out in the $NpT$ ensemble at 300~K using
a time step of 0.2 fs, and at standard pressure (1 atm) for a total
simulation time of 5 ns. For each cluster size $N\in [2, 60]$, 1000
configurations were extracted yielding a total of 59000
snapshots. Geometrically dependent distributed charges were obtained
using kernelized MDCM (kMDCM),\cite{MM.kmdcm:2024} by fitting to the
electrostatic potential of an ensemble of 180 water monomer
structures at the CCSD(T)/aug-cc-pVTZ level of theory. The structures
were defined by 20 evenly spaced angles between $84.45^{\circ}$ and
$120.45^{\circ}$, and OH bond lengths of 0.909, 0.959, and 1.009
\AA\/. In brief, the kMDCM model for water was
obtained\cite{MM.kmdcm:2024} by fitting a characteristic set of
optimized non-equilibrium charge models, using a Gaussian kernel-based
representation to describe anisotropic electrostatics which change
smoothly with molecular geometry. Further details on the fitting
procedure can be found in Ref. \citenum{MM.kmdcm:2024}.\\

\noindent
Reference data for fitting the Lennard-Jones parameters was obtained
using hybrid DFT
($\omega$B97X-V/def2-QZVP)\cite{mardirossian2014omegab97x}and
CCSD(T)-F12B/aug-cc-pVTZ-F12 calculations. As DFT calculations are
computationally much more efficient and require less memory, reference
data for cluster sizes (H$_2$O)$_2$ up to (H$_2$O)$_{60}$ were
computed. For the much more demanding CCSD(T) calculations, the
maximum cluster size was (H$_2$O)$_4$. For the present work,
$\omega$B97X-V/def2-QZVP was thoroughly bench-marked against
high-level CCSD(T)-F12/aug-cc-pVTZ level energies for (H$_2$O)$_2$ to
(H$_2$O)$_4$ (see Figure~\ref{sifig:dft_vs_ccf12_benchmark}). This
hybrid functional\cite{mardirossian2014omegab97x} was also found to be
particularly suitable for intermolecular and non-covalent
interactions.\cite{goerigk2017look} \\

\noindent
With the electrostatics handled through the kMDCM model the only
remaining contribution to the total energy within the concept of an
empirical energy function is the van der Waals energy. This
contribution was determined by fitting LJ-parameters to reference
cluster interaction energies $E_{\rm int}^{\rm ref} = E_{\rm
  cluster}^{\rm ref} - \sum_{i=1}^{N} E_{\rm monomer,i}^{\rm ref}$
containing $N$ monomers. Here, $E_{\rm cluster}^{\rm ref}$ is the
total energy of a cluster of given size $N$, and $E_{\rm
  monomer,i}^{\rm ref}$ is the energy of monomer $i$ in its actual
geometry which are all obtained either from $\omega$B97X-V/def2-QZVP
or CCSD(T)-F12B/aug-cc-pVTZ-F12 calculations. Likewise, $E_{\rm
  monomer,i}^{\rm model}$, $E_{\rm cluster}^{\rm model}$, and $E_{\rm
  int}^{\rm model}$, are the energies based on the NN-representation
of the bonded interactions, using the kMDCM model for the
electrostatics, and a particular choice of LJ-parameters. For each
cluster size $N$, at least 1000 different geometries were used. \\

\noindent
Starting from the total electronic energy, the internal monomer
energies (calculated using the neural network monomer potential) were
subtracted to obtain cluster interaction energies. Next, the
electrostatic contribution as obtained from the kMDCM model was
subtracted, leaving the remaining energy contribution to be fitted by
the LJ pair-potential.\cite{MM.ff:2024} The two models that arise from
using reference data at the $\omega$B97X-V/def2-QZVP or
CCSD(T)-F12B/aug-cc-pVTZ-F12 levels are referred to as \mdfts and
\mccs in the following. \\

\subsection{MD Simulations and Analysis}
MD simulations were run with CHARMM version c48.\cite{MM.charmm:2024}
The simulation system was a cubic box with 2000 water molecules and
periodic boundary conditions were applied. First, the system was
equilibrated at 300 K in the $NVT$ ensemble using the Nos\'e-Hoover
temperature bath\cite{Nose:1984} with a temperature window of 10
K. Following heating, equilibration in an isothermal-isobaric ($NpT$)
ensemble using a pressure bath at 1 atm coupled to a single
Nos\'e-Hoover temperature bath using an integration time step of
$\Delta t = 0.2$ fs was performed. All simulations were run with SHIFT
and SWITCH cutoff functions for nonbonded electrostatics and van der
Waals interactions, respectively.\cite{charmm:2009} The
switching-function parameters were $R_\textrm{on} = 10.0$ \AA\/ and
$R_\textrm{off} = 12.0$ \AA\/, respectively, for nonbonded van der
Waals interactions and a 12.0 \AA\/ cutoff was applied to the shifted
nonbonded electrostatics.  Production simulations were typically 2 ns
in length; however, self-diffusivities were calculated from repeat 10
ns simulations with different system sizes.\\

\noindent
The following quantities were determined from the MD simulations using
standard expressions, and were compared to experiment wherever
possible:\\

\noindent
{\bf Pure liquid density:} $\rho$ was calculated from $\langle \rho
\rangle = \frac{N_{\mathrm{res}}M}{N_{\mathrm{A}} \langle V \rangle}$
in the $NpT$ ensemble.\\

\noindent
{\bf Radial distribution functions:} $g_{\rm OO}$ was determined for
$0 < r_{\rm OO} < 10$ \AA\/ using $\Delta r_{\rm OO} = 0.1$~\AA\/
using VMD\cite{VMD} and applying periodic boundary conditions.\\

\noindent
{\bf Tetrahedral order parameter:} The parameter $q_{\text{tet}}$ is
defined as\cite{Duboue:2015}
\begin{equation}
    q_{\text{tet}} = 1 - \frac{3}{8} \sum_{j=1}^{3} \sum_{k=j+1}^{4}
    \left( \cos \psi_{ijk} + \frac{1}{3} \right)^2
\end{equation}
where $\psi_{ijk}$ are the angles between the oxygen atom of the
central water molecule $i$, and all pairs of oxygen atoms sampled from
the four nearest neighbor water molecules with indices $j$ and $k$. A
completely disordered structure corresponds to $q_{\text{tet}} = 0$
whereas for a perfect tetrahedral arrangement $q_{\text{tet}} = 1$.\\

\noindent
Experimental thermodynamic observables were also considered:\\

\noindent
{\bf Heat of vaporization:} $\Delta H_{\mathrm{vap.}}$ was determined
from
\begin{equation}
\Delta H_{\mathrm{vap.}}(T) = E^{\mathrm{potential}}_{\mathrm{gas}}(T)
- E^{\mathrm{potential}}_{\mathrm{liquid}}(T) + RT.
\label{eq:heatovap}
\end{equation}
where $R$ is the ideal gas constant,
$E^{\mathrm{potential}}_{\mathrm{gas}}(T)$ is the average potential
energy of a single water molecule in the gas phase, and
$E^{\mathrm{potential}}_{\mathrm{liquid}}(T)$ is defined as the
average total potential energy divided by the number of
molecules. Heating to and equilibration at 300 K was controlled
through the Nose-Hoover thermostat\cite{Nose:1984} whereas the
production run was carried out in the $NVE$ ensemble.\\

\noindent
{\bf Hydration Free Energy:} Hydration free energies $\Delta
G_\text{hyd}$ for the water in the bulk were computed from
thermodynamic integration. One water molecule was sampled in the gas
phase and in pure water. The condensed-phase simulations were carried
out in the $NpT$ ensemble with 2000 water molecules. Simulations with
39 evenly spaced coupling parameters $\lambda \in (0, 1)$ for
electrostatic and vdW interactions were performed. For each value of
$\lambda$ the simulation system was equilibrated for 50 ps after which
150 ps of production dynamics was carried out from which the hydration
free energy was accumulated according to
\begin{equation}
    \Delta G_\text{hyd} = \sum_{\lambda} \left[ \left(
      H^\text{elec}_\text{solv}(\lambda) -
      H^\text{elec}_\text{gas}(\lambda) \right) + \left(
      H^\text{vdW}_\text{solv}(\lambda) -
      H^\text{vdW}_\text{gas}(\lambda) \right) \right] \Delta \lambda
\end{equation}
The total hydration free energy was obtained from numerical
integration using the trapezoidal rule by weighting the variance of
the first and last window by 1/4 to compute
errors.\cite{burden:1997}\\

\noindent
{\bf Isothermal compressibility:} $\kappa$ was determined from
standard fluctuation formulae\cite{jenson:1998}
\begin{equation}
    \kappa = - \frac{1}{V} \Big( \frac{\partial H}{\partial T}
    \Big)_{N,P} = \frac{1}{k_BT\langle V \rangle} (\langle V^2\rangle
    - \langle V \rangle^2)
\end{equation}
and standard deviations were calculated using the Circular Block
Bootstrap method\cite{Dudek:2015} for time series data using the
volume $V$ of the simulation box in $NpT$ simulations.\\

\noindent
{\bf Dielectric constant:} From the fluctuations of the total dipole
moment of the box, $M$, the dielectric constant was obtained according
to\cite{Neumann:1983}
\begin{equation}\label{eq:dielectric_const}
    \epsilon = 1 + \frac{4\pi}{3\epsilon_0 V k_{B}T} [ \langle M^{2}
      \rangle - \langle M \rangle^{2}].
\end{equation}
In Equation~\ref{eq:dielectric_const}, $\epsilon_0$ is the vacuum
permittivity and the box volume $V$ was constant as simulations were
performed in the $NVT$ ensemble.\\

\noindent
Finally, two transport properties were considered:\\

\noindent
{\bf Self-diffusion coefficient:} $D_{\mathrm{PBC}}$ was computed from
the mean squared displacement of all oxygen atoms using the Einstein
relation
\begin{equation}
    D_{\mathrm{PBC}} = \lim_{t \longrightarrow \infty} \frac{1}{6t}
    \langle | r(t) - r(0)|^{2} \rangle.
\end{equation}
For this, the trajectories must be ``unwrapped" so that the particles
do not reenter the central image cell when passing through periodic
boundaries but rather diffuse naturally in all dimensions; this was
achieved using the PBC unwrap tool available in VMD.\cite{VMD}
Simulations were performed in the $NVT$ ensemble. Additional 2 ns of
$NVE$ simulations were performed to confirm the average mean squared
displacement (MSD) was consistent and to quantify energy
conservation.\\

\noindent
Diffusion coefficients estimated from MD simulation require
corrections\cite{kremer:1993,Yeh:2004,MM.hb:2018} due to finite size
effects arising from statistical `self-interactions' in the long time
limit. For a cubic simulation box of length \( L \), this is given by:
\begin{equation}
    D_0 = D_{\text{PBC}} + \frac{\xi \, k_B T}{6 \pi \eta L}
    \label{eq:Dcorr}
\end{equation}
where \( D_{\text{PBC}} \) is the diffusion coefficient calculated in
the simulation, \( k_B \) is the Boltzmann constant, \( T \) the
absolute temperature, and \( \eta \) the shear viscosity of the
solvent, and the Ewald self-correction constant $\xi \simeq
2.837297$ for cubic lattices.\cite{Yeh:2004}  \\

\noindent
{\bf Angular reorientation times:} Orientational correlation functions
were calculated using
\begin{equation}
    C_2(t) = \left\langle P_2 \left( \vec{e}(0) \cdot \vec{e}(t)
    \right) \right\rangle.
\end{equation}
Here, $\vec{e}$ represents a unit vector along one of the OH bonds of
a water molecule, $P_2$ is the second-order Legendre polynomial, and
the bracket indicates an ensemble average over all OH bonds at a given
time $t$. The average water orientational relaxation time $\tau_2$ was
determined by fitting the long-time decay of the corresponding
orientational correlation functions to a single exponential
function. Results were averaged over 40 non-overlapping time series
from 2 independent trajectories, with a time resolution of 0.2 fs
simulated over 20 ps in the $NVT$ ensemble.\\

\section{Results and Discussion}
The present section evaluates the performance of the developed
interaction model. First, the accuracy of the intramolecular potential
is analyzed to ensure it accurately captures the relevant molecular
energies and forces. Following this, the inter- and intramolecular
models are combined to assess the performance of this ML-based energy
function on clusters and condensed phase properties.\\

\subsection{Performance of the Intramolecular NN PES}
The intramolecular NN PES for water is sought to accurately reproduce
\textit{ab initio} CCSD(T)-F12B/aug-cc-pVTZ-F12 energies and forces in
a highly efficient manner. Once the PES is trained, the need to
repeatedly solve the electronic structure problem is circumvented by
learning the mapping $f$ from interatomic distances to potential
energies ($f:\{r_i\} \rightarrow E$) directly. While on average the
\textit{ab initio} calculations take 4.5 min, the evaluation time for
the NN PES (\textit{i.e.}, one $E$ and $\bm{F}$ evaluation) takes
0.85~ms (0.0026~ms) when a Python (Fortran) implementation is
used. This is a significant speed-up of several orders of magnitude
and the relative timings are $\sim 10^8/10^2/1$ for {\it ab
  initio}/Python/Fortran, respectively.\\

\begin{table}[h]
\begin{tabular}{ll}\toprule
\textbf{kcal/mol(/\AA)}      & \textbf{NN} \\\midrule
MAE($E$) & 0.0073\\
RMSE($E$) & 0.0141\\
MAE($\bf{F}$) & 0.0758 \\
RMSE($\bf{F}$) & 0.1532 \\
$1-R^2$ & 5.3E-06 \\\bottomrule
\end{tabular}
\caption{Mean absolute (MAE) and root mean squared errors (RMSEs) of
  energies and forces on a hold-out test set containing 366 random
  water monomer structures.}
\label{tab:outofsample_errors}
\end{table}

\begin{figure}[h!]
\centering
\includegraphics[width=0.8\textwidth]{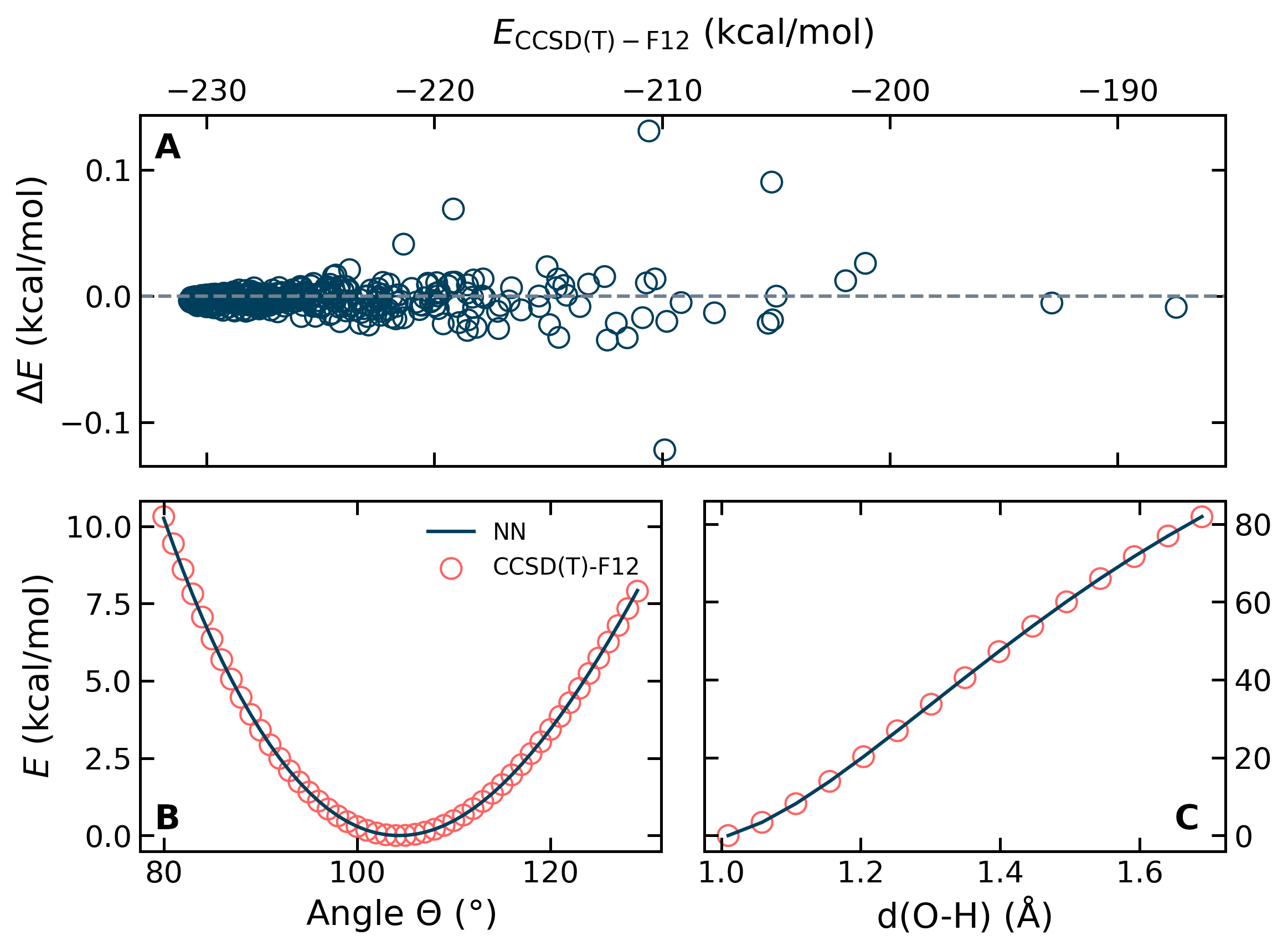}
\caption{Performance of the NN PES \textbf{A:} Prediction error
  $\Delta E = E_{\rm CCSD(T)-F12} - E_{\rm NN}$ on a hold-out test set
  containing 366 random water structures. The predictions are all
  within $\sim 0.1$~kcal/mol and have a high coefficient of
  determination ($1 - R^2 = 5.3E-06$). The potential energy of a water
  molecule as a function of the bending angle $\Theta$ (\textbf{B})
  and of the O--H bond length (\textbf{C}). For \textbf{B} and
  \textbf{C} the energy is given with respect to the minimized
  structure of water and is compared to direct CCSD(T)-F12 energies.}
\label{fig:ffnet_corr_water_cc}
\end{figure}

\noindent
Ideally, this gain in computational efficiency does not come at the
expense of accuracy. The energetics of the NN PES are assessed and
compared to their \textit{ab initio} reference in
Figure~\ref{fig:ffnet_corr_water_cc}.  In panel A, the prediction
error of the NN PES, $\Delta E = E_{\rm CCSD(T)-F12} - E_{\rm NN}$, is
shown for a hold-out test set containing 366 random water
conformations. It shows that the prediction errors are within $\sim
0.1$~kcal/mol of direct \textit{ab initio} energies and, hence, are
far below ``chemical accuracy'' (an empirical threshold beyond which
qualitative characterization of a many-body system is ought to be
possible\cite{wang2024design}). The corresponding mean absolute and
root mean squared errors (MAE and RMSE, respectively) for the hold-out
test set are given in Table~\ref{tab:outofsample_errors}. Panels B and
C of Figure~\ref{fig:ffnet_corr_water_cc} report the potential energy
of the NN PES as a function of the bending angle and an O--H bond
length, respectively. Both scans reproduce the \textit{ab initio}
energy with high fidelity and show maximum differences of $\sim
0.05$~kcal/mol for strong molecular distortions with $\theta \sim
80^\circ$ and $d$(O-H) $\sim 1.5$\AA, respectively.\\

\noindent
Finally, the accuracy of the atomic forces, which are required for
faithful and stable MD simulations can be evaluated. Besides
inspecting averaged quantities, such as MAE($\bm{F}$) determined on a
test set (see Table~\ref{tab:outofsample_errors}) a measure for the
reliability of the forces is the energy conservation within a
finite-$T$ MD simulation. For a gas-phase $NVE$ simulation ($\Delta t
= 0.25$~fs, 800000 steps, random momenta drawn from Maxwell-Boltzmann
distribution corresponding to 300~K) the energy is conserved to within
a standard deviation of $6\cdot 10^{-4}$~kcal/mol and a maximum
fluctuation of $10^{-3}$~kcal/mol.\\

\noindent
Harmonic and anharmonic frequencies determined at the minimum water
configuration can serve as another meaningful proxy for the accuracy
of the forces and higher order derivatives. Vibrational frequencies
depend on the curvature of the PES and are given in
Table~\ref{tab:harm_freq_water_ccf12}.  Harmonic frequencies for the
NN PES reproduce their \textit{ab initio} reference with a
MAE($\omega$) of 1.5~cm$^{-1}$ which emphasizes the high quality of
the PES with respect to the reference data. Spectroscopically, it is
interesting to assess and compare the computed vibrational frequencies
of a gas-phase water molecule with experiments. Including (some)
anharmonicity into the vibrational modes is straightforward to achieve
with MD simulations, however, the vibrations typically remain
blue-shifted with respect to experiment.\cite{suhm:2020,MM.fad:2022}
This is particularly evident for the high-frequency stretching modes
(\textit{e.g.}, O--H) which are $\sim 150$~cm$^{-1}$ higher than the
measured frequencies. This is a limitation of classical dynamics
simulations and their zero-point energy
leakage\cite{guo1996analysis}. Therefore, anharmonicity is better
accounted for by using, \textit{e.g.}, second-order vibrational
perturbation theory (VPT2)\cite{piccardo2015generalized}, which was
recently combined with ML models.\cite{MM.tl:2021} The VPT2
frequencies determined on the present water PES are given in
Table~\ref{tab:harm_freq_water_ccf12} and their comparison to
gas-phase measurements yields a MAE($\nu$) of 3.3~cm$^{-1}$ which
underscores the high quality of the ``bonded'' part of the interaction
potential compared with experiment.\\

\begin{table}[h]
\begin{tabular}{lccccccc}\toprule
     & \multicolumn{4}{c}{harmonic} & \multicolumn{3}{c}{anharmonic}\\\cmidrule(lr){2-5}\cmidrule(lr){6-8}
  & NN & NN$_{\rm MD}$(g)& NN$_{\rm MD}$(l) & {\it ab initio}(g) & NN$_{\rm VPT2}$(g)  & Exp.\cite{fraley1969high,benedict1956rotation}(g) & Exp.\cite{bertie:1996}(l)\\
  \midrule
$\nu_2$  & 1649.2   & 1646.6   & 1750  & 1648.4  & 1599.4 & 1594.6 & $\sim 1650$\\
$\nu_1$  & 3828.8   & 3809.6  & 3780   & 3828.6  & 3655.5 & 3657.1 & $\sim 3400$ \\
  $\nu_3$  & 3935.3   & 3919.2  & 3870   & 3938.8  & 3752.4 & 3756.0 & $\sim 3400$\\
  \bottomrule
\end{tabular}
\caption{Vibrational frequencies (in cm$^{-1}$) of water: The harmonic
  frequencies from diagonalizing the mass-weighted Hessian matrix are
  compared with \textit{ab initio} CCSD(T)-F12B/aug-cc-pVTZ-F12
  frequencies and with those determined from gas-phase MD
  simulations. Anharmonicity is included by using the VPT2 method as
  implemented in Gaussian 16\cite{g16}. These are compared to gas-(g)
  and liquid-(l) phase experiments.}
\label{tab:harm_freq_water_ccf12}
\end{table}

\subsection{Fitting and Performance of the Intermolecular Model}
Fitting of the LJ-parameters was carried out for 1000 configurations
of clusters binned by size $N$. Sampling of the conformations and
cluster sizes favoured smaller cluster sizes. For the DFT-based
reference data set, 1000 structures were drawn from a set in which the
maximum cluster size increased from $N \in [2,59]$ in steps of 1
whereas for the CCSD(T) data set the 1000 structures were drawn from
the entire data set $N \in [2,4]$. Optimized Lennard-Jones parameters
were obtained from 200 random initial values sampled around $\pm 10$\%
of the original CHARMM TIP3 parameters\cite{brooks:2004} using the
Nelder-Mead algorithm\cite{nelder:1965} and $\mathcal{L} = | E_{\rm
  inter}^{\rm ref} - E_{\rm inter}^{\rm model} |_2$ as the loss
function. Three repeats of this procedure were carried out, resulting
in 180 sets of structures for each of which 200 independent
optimizations of the parameters were carried out. This leads to a
total of 36000 parameter combinations of which the 300 best performing
combinations were considered for further processing.\\

\begin{figure}[h!]
    \centering
    \includegraphics[width=1.0\linewidth]{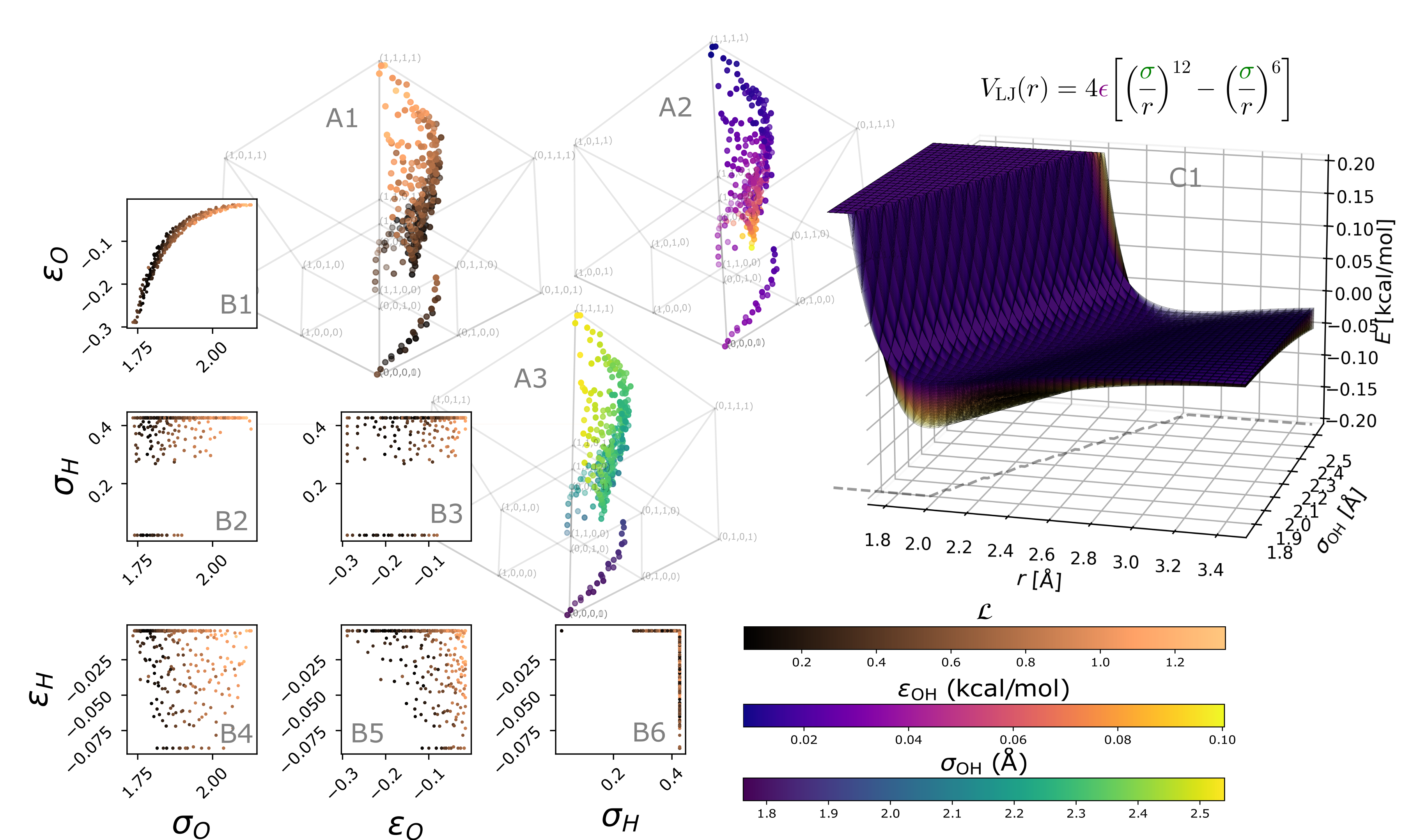}
    \caption{The LJ-Fit for \mcc. Panel A1: Projections of the four
      fitted LJ-parameters using Schlegel's hypercube representation
      (see text), with vertices annotated as 0 (minima) or 1 (maxima)
      in the transformed ($\sigma_{\rm O}$, $\sigma_{\rm H}$,
      $\epsilon_{\rm O}$, $\epsilon_{\rm H}$) coordinate system. The
      three color scales are for $\sigma_{\rm OH} = (\sigma_{\rm
        O}+\sigma_{\rm H})/2$, the values of the loss $\mathcal{L} =
      \Delta \rho / 2 + \Delta(\Delta H)$, and $\epsilon_{\rm OH} =
      \epsilon_{\rm O} \epsilon_{\rm H}$, respectively. Panel B:
      Correlations among the fitted LJ-parameters: $\sigma_{\rm O}$,
      $\epsilon_{\rm OH}$, and $\sigma_{\rm H}$, $\epsilon_{\rm H}$.
      Panel C1: The LJ-potential as a function of distance and radii
      $\sigma$ for the OH combined parameters with slices colored by
      $\epsilon_{\rm OH}$ over the range of parameters tested; the
      position of $r_{\rm min}$ is shown as a gray dashed line.}
    \label{fig:params1}
\end{figure}

\begin{figure}[h!]
    \centering
    \includegraphics[width=0.995\linewidth]{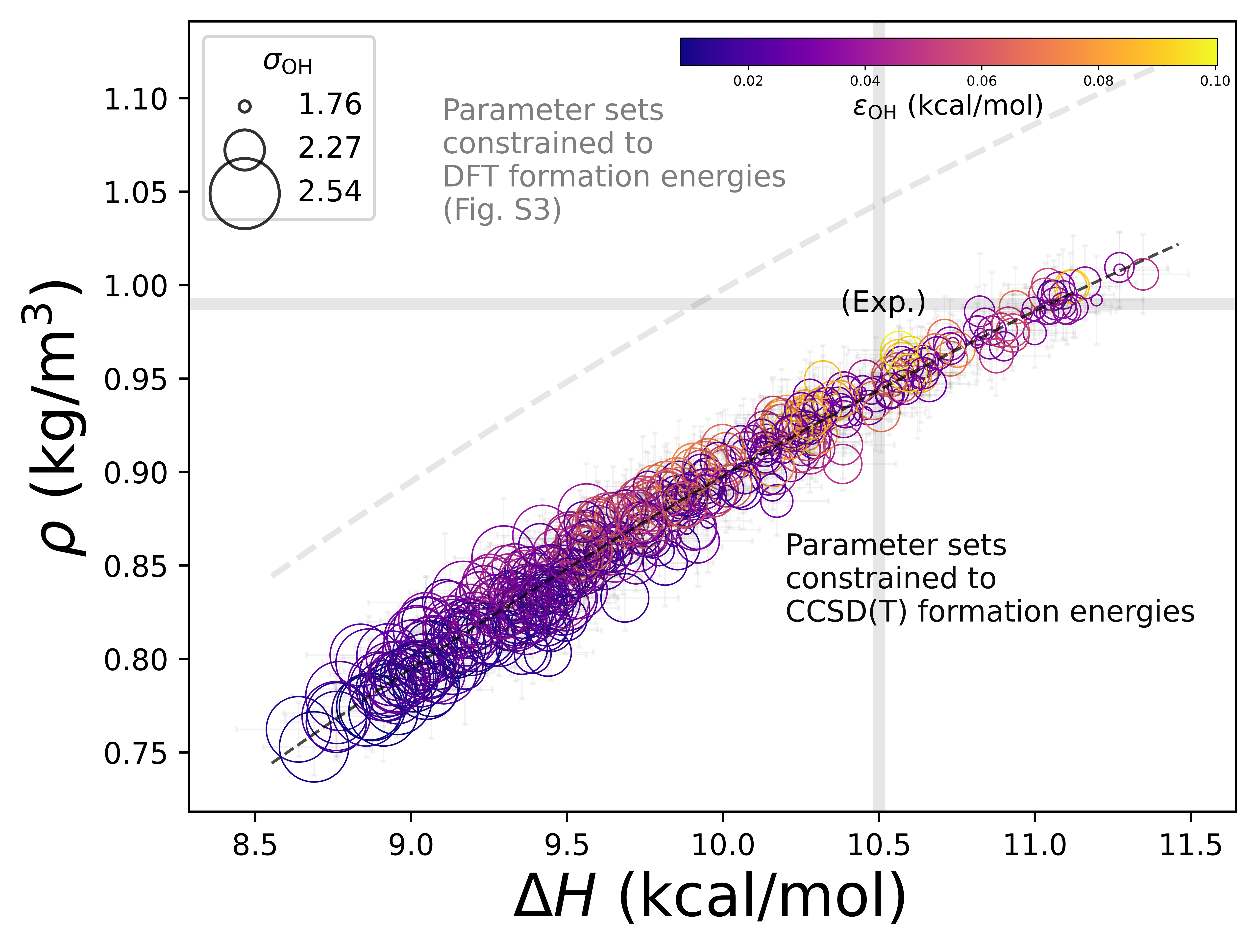}
    \caption{Simulated density and enthalpy of vaporization of ambient
      water using the candidate \mccs Lennard-Jones parameters. The
      size of each data point corresponds $\sigma_{\rm OH} =
      (\sigma_{\rm O}+\sigma_{\rm H})/2$ parameter, while the color
      represents the corresponding $\sqrt{\epsilon_O\epsilon_ H}$
      value. The dashed grey line plots the trends observed using the
      \mdfts model.}
    \label{fig:params2}
\end{figure}

\noindent
All 300 of the top performing parameter sets obtained from fitting
yield a final RMSE between reference energies and fitted energies
better than 1 kcal/mol.  These parameters covered a wide ranges of
values (see Figure \ref{fig:params1} panels A1, A2, A3 and B for the
\mccs model), although several of the parameter combinations are
correlated.  For example, the van der Waals well depth and size of the
oxygen atom is correlated over the entire range of parameter values
(panel A2), although different combinations yield lower or higher
values of the loss function as indicated by the color code in panel
A2. A Schlegel hypercube representation allows to present
4-dimensional data in 3 dimensions by nesting a cube inside the
existing 3D Euclidean space \textit{via} a set of constrained linear
equations;\cite{schlegel:1886} this was done for $(\sigma_{\rm O},
\sigma_{\rm H}, \epsilon_{\rm O}, \epsilon_{\rm H})$, see Figure
\ref{fig:params1} panels A1-3. Using this projection isolates two
populations, a majority and a population where $\sigma{\rm H}$ is at a
minimum (likely a local minimum of $\mathcal{L}$ found during fitting)
and $\epsilon$ is decreased to compensate.  The trade-off between
$\epsilon$ and $\sigma$ is shown graphically in Figure
\ref{fig:params1}C1 where the comparable 'slices' of the potential as
a function of $r$ can be chosen by varying $\epsilon$ and $\sigma$.\\

\begin{table}[!t]
\centering
\caption{Training reference data: Density and enthalpy of vaporization
  for various water models.\cite{MM.dcm:2020,Fadda:2011} Density and
  enthalpy of vaporization were used to select LJ parameters for the
  kMDCM-NN models which were predicted in reference to DFT or CCSD(T)
  interaction energies.}
\begin{tabular}{cllc}
\toprule
& Model & $\rho$ (g/ml) & $\Delta H_{\mathrm{vap.}}$ (kcal/mol) \\
\midrule
 & Experiment & ~0.996 & ~10.51 $\pm$ 0.01 \\\midrule
\multirow{2}{*}{kMDCM} & \mdft & ~1.002 & ~10.3 $\pm$ 0.1 \\
& \mcc & ~0.997 & ~10.6 $\pm$ 0.1 \\
\midrule
\multirow{3}{*}{TIP} & TIP3P\cite{MM.dcm:2020} & ~1.027 & ~11.04 \\
& TIP4P\cite{MM.dcm:2020} & ~1.009 & ~10.6 \\
& TIP4P/2005\cite{abascal:2005} & ~1.009 & ~11.99 \\
& TIP5P\cite{MM.dcm:2020} & ~0.985 & ~10.73 \\
\midrule
Polarizable & iAMOEBA\cite{Wang:2011} & 0.997 & 10.5 \\
\bottomrule
\end{tabular}
\label{tab:prop-train}
\end{table}

\noindent
Evidently, fitting to cluster interaction energies does not
unambiguously distinguish between the various solutions of the
optimization problem. For further model selection out of the 300
fitted \mdfts and \mccs models, the pure liquid density ($\rho$) and
enthalpy of vaporization ($\Delta H_{\rm vap}$) were determined from
MD simulations using each of the 300 models and comparing with
measured $[\rho,\Delta H_{\rm vap}]$ values as is routinely done in
force field development.\cite{cgenff:2010} These simulations used the
NN model for the bonded terms together with the kMDCM electrostatics
and each of the fitted parameter sets and were carried out at standard
conditions (298~K, 1~atm) for 1000 water molecules in a 35 \AA\/ cubic
box with periodic boundary conditions. Figure \ref{fig:params2}
reports the two thermodynamic variables which are found to be well
correlated with one another, as expected. The experimentally reported
values $[\rho_{\rm Exp.}, \Delta H_{\rm Exp.}]$ are indicated at the
crossing point. Broadly speaking, models with smaller oxygen van der
Waals radii ($\sigma_{\rm O}$) are more consistent with experiments
whereas with increasing $\sigma_{\rm O}$ model performance
deteriorates. To determine the preferred (``best performing'') model,
the loss function $\mathcal{L} = 0.5$ MAE$_{\rho}$ + MAE$_{\Delta H}$
was used. This yields a model which performs for $\rho$ and $\Delta H$
as reported in Table \ref{tab:prop-train} and the parameters are
summarized in Table \ref{sitab:atom_params}. Henceforth only these
models are used for \mdfts and \mcc, respectively.\\

\subsection{Gas-Phase Benchmarks for Water Hexamer Clusters}
\begin{figure}[h!]
    \centering
    \includegraphics[width=0.9\linewidth]{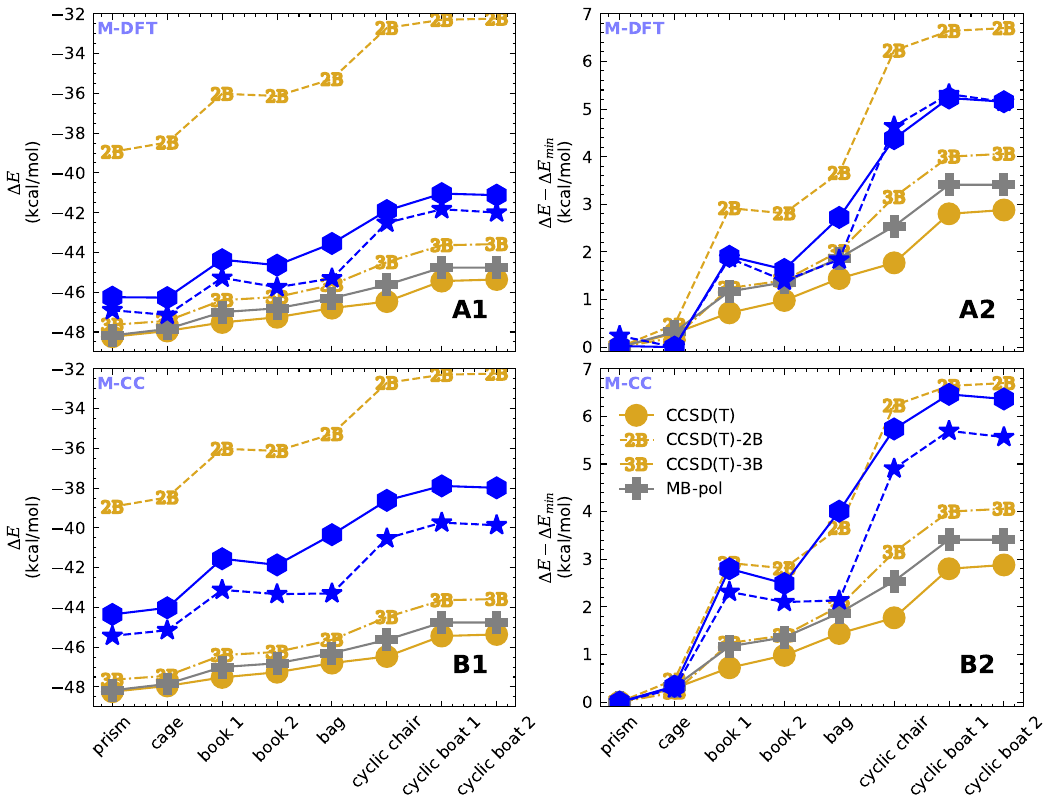}
    \caption{Total interaction and differential stabilization energies
      for water hexamer from using models \mdfts (top) and \mccs
      (bottom), together with results from CCSD(T)/CBS reference
      data\cite{reddy:2016} (gold) and MB-pol (grey). The geometries
      from \mdfts and \mccs were optimized using steepest descent
      (10000 steps, initial step size of 0.0002 \AA) in CHARMM (blue,
      dashed) and showed small conformational changes (see
      Fig. \ref{fig:hexamers_freq}). Panels (A1/B1) Interaction
      energies of the low-lying isomers of the (H$_2$O)$_6$
      clusters. Panels (A2/B2) Interaction energies relative to the
      lowest energy conformer.  RMSE values between reference CCSD(T)
      and model energies for the original/optimized geometries were
      (\mdft) 3.4/2.6 and (\mcc) 6.2/4.5 kcal/mol.}
    \label{fig:hexamers}
\end{figure}

Next, the fitted models \mdfts and \mccs were used to probe finer
details of the intermolecular interactions. This is, for example,
encoded in the structures, relative interaction energies and normal
mode frequencies of water hexamers (H$_2$O)$_6$, see Figure
\ref{fig:hexamers}). For this system, high-quality benchmark
calculations are available including 2-body, 2- and 3-body
contributions, or at the CCSD(T)/CBS limit.\cite{reddy:2016}\\

\noindent
The structures of all 8 isomers (prism to cyclic boat 2) were
optimized within CHARMM using \mdfts and \mccs and the formation
energies with respect to the separated monomers are reported in
Figures \ref{fig:hexamers}A1 and B1. All optimized structures compare
very favourably with those at the CCSD(T)/CBS level with overall ${\rm
  RMSD} < 0.1$ \AA\/. Compared with CCSD(T)/CBS the total interaction
energies from \mdfts and \mccs are smaller in magnitude by 2 to 4
kcal/mol and 4 to 8 kcal/mol, respectively. In terms of stabilization
energy relative to the lowest-energy conformer, the differences reduce
to 2 kcal/mol and 3 kcal/mol, respectively. Most notably, \mccs
correctly predicts the prism structure to be most stable. This is not
the case for \mdfts for which the cage is slightly more stable. MB-pol
(grey), which used the hexamer energies in the fitting, is clearly
superior in estimating the interaction/stabilization energies to the
models considered here. In comparison, CHARMM's TIP3 water
model\cite{brooks:2004} almost completely reverses the ranking of the
hexamers, see Figure \ref{sifig:hexamers_tip3}. Note that models such
as iAMOEBA,\cite{wang:2013} q-AQUA,\cite{yu:2022} or
MB-pol\cite{Zhu:2023} include the hexamer data during
parametrization. Consequently, the errors compared with CCSD(T)/CBS
data is necessarily smaller (mean absolute errors of 0.5 kcal/mol or
below for all models).\\

\noindent
Considering the reference data that includes 2B-contributions at the
CCSD(T) level it is found that \mdfts and \mccs follow these trends
rather closely. When further including 3B-contributions the
interaction energy (CCSD(T)-3B) changes considerably more gradually
compared with (CCSD(T)-2B). This implies that \mdfts and \mccs are
rather successful in capturing the two-body contributions at the
CCSD(T)-2B level. The improved `two-body quality' of this parameter
set may be explained by the over representation of dimers in the
training set, which make up over 90\% of the data. On the other hand,
further accounting for 3B interactions in the parametrization is
expected to boost their performance.\\

\begin{figure}[h!]
    \centering
    \includegraphics[width=1.0\linewidth]{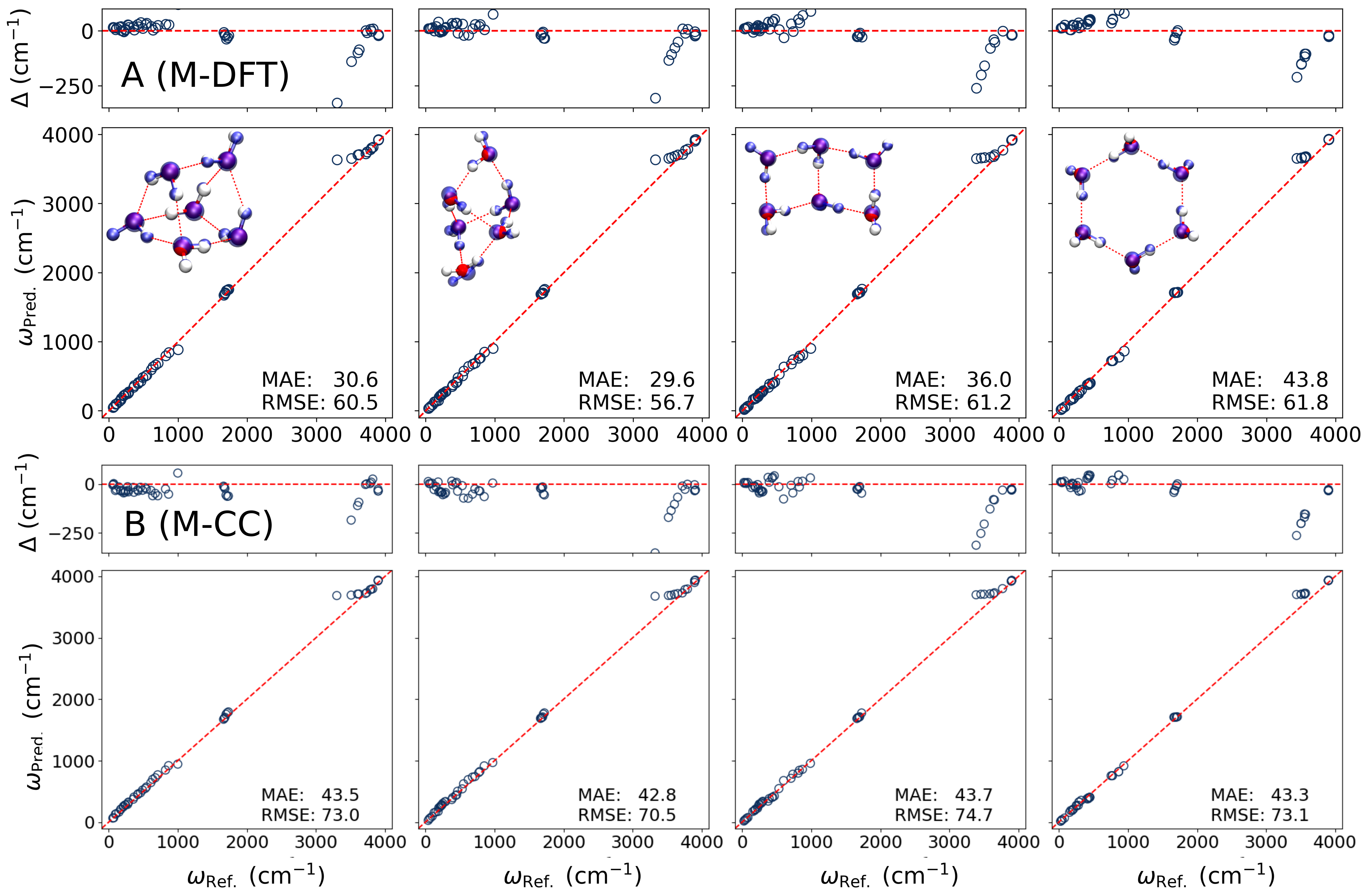}
    \caption{Comparisons of the harmonic frequencies of four low-lying
      hexamer minima (from left to right: prism, cage, book 1 and
      cyclic chair as obtained from \mdfts~(A) or \mccs~(B) with
      Reference~\citenum{howard:2015}. The top panels show the signed
      error of the prediction (\textit{i.e.}, $\Delta = \omega_{\rm
        Pred.} - \omega_{\rm Ref.}$). The insets show the difference
      between the optimized structures and the structures reported in
      Reference~\citenum{howard:2015} (blue).}
    \label{fig:hexamers_freq}
\end{figure}

\noindent
Based on the optimized structure of the prism, cage, book1 and cyclic
chair structures, normal mode calculations using \mdfts and \mccs were
carried out. For \mdfts the MAE between the reference CCSD(T)/CBS
calculations\cite{howard:2015} ranged from 30 to 44 cm$^{-1}$ for the
4 structures considered, see top row in Figure
\ref{fig:hexamers_freq}. It is also noted that the low-frequency
(framework) modes are better captured, whereas in the region of the
OH-stretch vibration, the differences can be larger. Specifically, the
lowest-frequency OH-stretch modes are shifted too much to the blue (by
up to 250 cm$^{-1}$) compared with the reference CCSD(T)
frequencies. For \mccs the correlation between model predictions and
reference harmonic frequencies follows similar patterns as for \mdfts
but the mean average errors are larger by up to $\sim 15$ cm$^{-1}$,
depending on the structure considered.\\

\noindent
Using \mdfts the sublimation energy of clusters containing 2 up to 60
monomers was determined by fitting the cluster decomposition energy as
a function of inverse system size and extrapolating to the limit of
infinite system size. This yielded $\Delta H_{\rm sublim}^{\rm comp}
\sim 10.2 \pm 0.2$ kcal/mol, see Figure \ref{sifig:sizeDe}C, compared
with a measured value of 10.2 kcal/mol.\cite{Feistel:2007} A similar
result was reported for the q-AQUA model (10.2 kcal/mol within
uncertainty).\cite{Bowman:2024} \

\subsection{Condensed Phase Simulations}
{\bf Density and Structural Properties:} Parametrizing a water model
to reproduce the bulk density of a liquid at standard conditions is
routine and usually straightforward despite the limited number of
fitting parameters, namely the four LJ parameters for oxygen and
hydrogen. The \mdfts and \mccs models reproduce the experimental bulk
density ($\rho$) of liquid water within 0.01 g/ml, similar to the TIP
water models, with all predictions falling within approximately 0.03
g/ml of the measured value, see Table \ref{tab:prop-train}. Variations
in reported density across computational studies often arise from
differences in the treatment of long-range electrostatics and choice
of cut-off distances. The density slightly increases for larger
nonbonded cut-off values. As shown in Figure \ref{fig:params2},
simulated densities and heats of vaporization consistent with
experiments can be obtained from a wide range of combinations of van
der Waals parameters ($r_{\rm min}$ (point size) and $\epsilon$ (point
color)). \\

\noindent
For MD simulations of hydrated species such as organic solutes or
proteins the bonds involving H-atoms are often constrained for
computational efficiency. Constraining the OH bond lengths using
SHAKE\cite{shake:1977} and a usual time step of $\Delta t = 1$ fs
afforded stable simulations. With the same nonbonded cutoffs as used
in the flexible simulations $\rho_{\rm calc} = 0.999$ kg/m$^3$ was
obtained compared with $\rho_{\rm expt.} = 0.998$ kg/m$^3$. This
result is still encouraging, as it suggests the bulk structural
features are consistent regardless of flexible or constrained water,
allowing practitioners to change between the two approaches depending
on the applications. This further increases applicability of the
present model.\\

\begin{figure}[h!]
\centering
\includegraphics[width=0.7\textwidth]{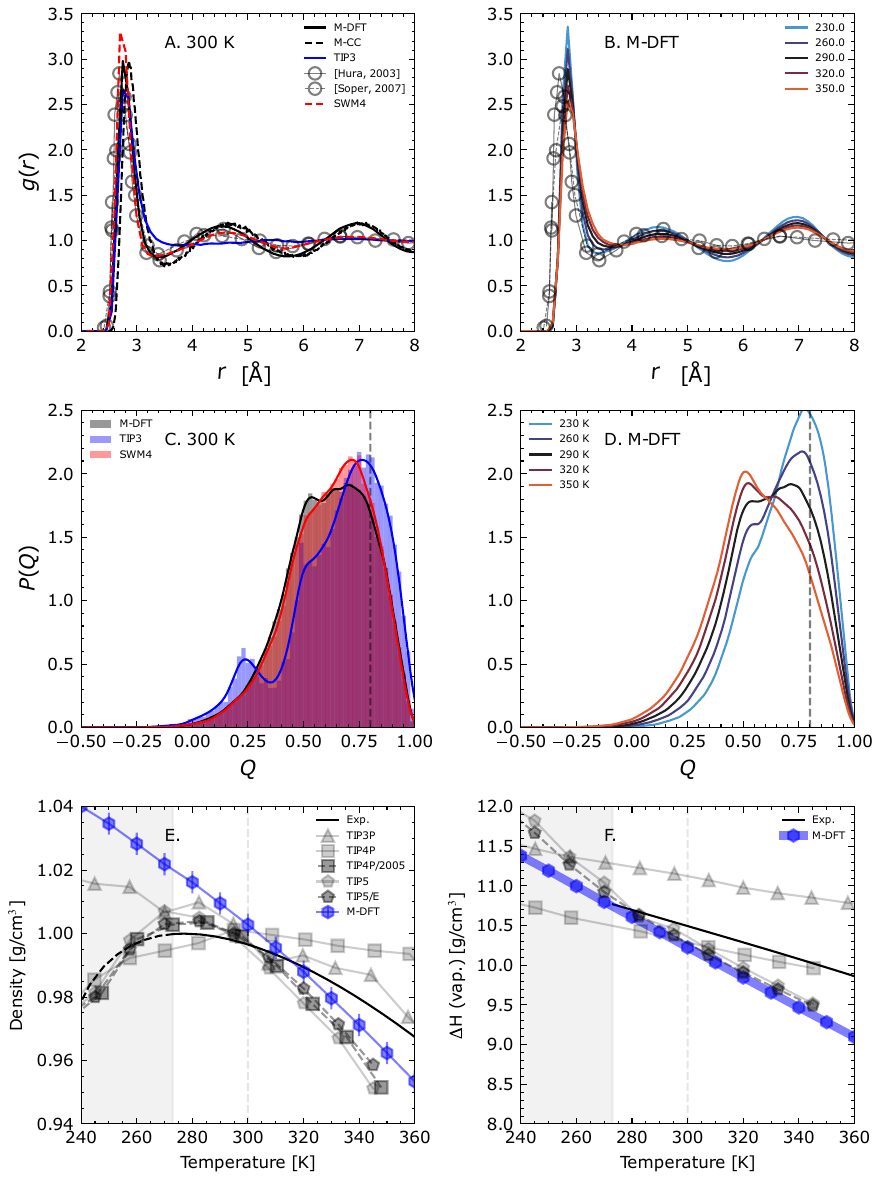}
\caption{Structural features of the bulk liquid water: (A) radial
  distribution functions $g(r)$ of \mdfts and \mcc, and (C)
  tetrahedral order parameter $P(Q)$ at 300 K of \mdfts, TIP3, and
  SWM4 potentials.  Temperature dependence of (C) $g(r)$ (D) $Q$ (E)
  density $\rho$, (F) heat of vaporization $\Delta H_{vap.}$ of \mdfts
  compared against literature values and experiment.}
\label{fig:tempDep}
\end{figure}

\noindent
Figure \ref{fig:tempDep} reports the O-O radial, and tetrahedrality
parameter, distributions (panels A to D), as well as results for the
pure liquid density (panel E), and the heat of vaporization (panel F,
discussed further below) as a function of temperature. All these
results correspond to performance on a ``test set of observables'' as
the \mdfts and \mccs models were only based on \textit{ab initio}
electronic energies and $[\rho, \Delta H_{\rm vap}]$ at 300 K.\\

\noindent
A measurable and useful structure-related property for validation is
the O--O pair radial distribution function (RDF) $g_{\rm
  OO}(r)$. Results from MD simulations in the $NpT$ ensemble using
models \mdfts and \mccs together with the TIP3P
model\cite{brooks:2004} implemented in CHARMM are reported in Figures
\ref{fig:tempDep}A and B. With the \mdfts model the positions of the
onset and first maximum of $g_{\rm OO}(r)$ are shifted by 0.1 \AA\/ to
larger separations but the peak height of the first peak is correctly
described. The first solvation shell is correctly located at $r_{\rm
  OO} \sim 4.5$ \AA\/, also matching the experimentally reported peak
height. However, the third solvation shell is over-structured in
comparison with experiment. As has been previously noted with the TIP3
water model, $g_{\rm OO}(r)$ is noticeably flat after the first peak
at $\sim 2.8$ \AA~ corresponding to the first solvation
shell.\cite{Wade:2018} As a further comparison, $g_{\rm OO}(r)$ from a
10 ns trajectory using the polarizable 4-point SWM4-HLJ water
model\cite{lamoureux:2006} was determined. This simulation had used a
water box containing 640 water molecules. The SWM4-HLJ model provides
the best agreement with the available experimental RDFs, with the
exception of the height of the first solvation shell which differs by
approximately 10\%.\\

\noindent
The distributed charge model kMDCM to describe conformationally
flexible electrostatics is based on a six-charge water model with two
`lone pair' sites on the oxygen forming a quasi-tetrahedral shape,
similar to the TIP5 water model.  There is evidence to suggest that
the pronounced tetrahedral geometry of TIP5 leads to over-structuring
of $g(r)$,\cite{Camisasca:2019} similar to observations for the kMDCM
model. Models such as q-AQUA\cite{qu2023interfacing} and
MB-Pol\cite{zhu2023mb} also include a distributed charge
representation to capture the first few multipole moments of the
electrostatic distribution of the water monomer and qualitatively
reproduce $g(r)$ within the model-based, and experimental,
uncertainties associated with obtaining such radial density
profiles. Interestingly, the $g(r)$ from the 1- and 2-body terms of
the successful q-AQUA model considerably misaligned with the measured
radial distribution function which is not what was found for \mdfts
and \mccs which are also 1- and 2-body models. This suggests that
including higher-order terms in the models discussed here will
considerably improve the performance.\\

\noindent
The tetrahedral order parameter $Q$ considered next provides a measure
for the ordering of water molecules inside the first solvation
shell.\cite{errington:2001,Duboue:2015} Notable comparisons include a
systematic shift to higher ordering (or more ice-like) structures for
TIP3 in comparison to \mdfts and SWM4 (Figure \ref{fig:tempDep}C).
Like the radial distribution functions, no prominent differences were
observed in the distributions of $Q$ for \mccs and \mdfts and the
former is excluded from Figure \ref{fig:tempDep}C for clarity. The
shoulder at $Q = 0.5$ was also observed in simulations using MB-Pol
and is temperature dependent, diminishing at lower
temperatures;\cite{Zhu:2023} it corresponds to transient reordering of
the hydrogen bond network. The relative probabilities between the
maximum at $Q = 0.7$ and the aforementioned shoulder are much closer
for kMDCM than the SWM4 model. The temperature dependence of the O-O
RDF and tetrahedral order parameter aid in illustrating these subtle
differences and are shown in Figures \ref{fig:tempDep}B and D.
Increased ordering of the hydration shells are in qualitative
agreement with wide angle x-ray scattering experiments of supercooled
water down. Perhaps fortuitously, the relative decrease in peak
heights between 230 K to 260 K is roughly 15\% (3.4 to 3.1) which is
comparable to the experimental change which reports a difference of
3.2 to 2.7 (arbitrary units).\cite{Pathak2019} While $P(Q)$ is not
readily amenable to measurements, Raman spectroscopy provides a
measure of the local electric field, and correlates with the expected
change in hydrogen-bond strength which decreases as a function of
temperature, similar to $Q$.\cite{Duboue:2015}\\

\noindent
With the TIP3 model $P(Q)$ is higher at low $Q$ values, suggesting a
tendency towards a collapse of the structure in the hydrogen bond
network, which is also consistent with the observed flatness in the
O-O RDF. By studying the temperature dependence of these properties
when using \mdft, the peak positions of the hydration shells are more
apparent, and are in agreement with the experimental values of 2.8
\AA\/ and 4.5 \AA\/ for the first two shells, although the third
solvation shell is further away at 7.0 \AA\/ in comparison to 6.8
\AA\/ as determined through analysis of the scattering pattern seen in
wide angle X-ray scattering experiments of supercooled water down,
and, despite limitations in reproducing the `freezing' behaviour due to
classical treatment of the liquid\cite{wang:2013,Ruiz:2018} increased
ordering of the hydration shells and tetrahedrality of the liquid at
low temperatures can provide meaningful interpretations regarding the
behaviour of the bulk environment.\\

\noindent
The temperature dependence of the bulk liquid density is shown in
Figure \ref{fig:tempDep}E, where decreases in density with respect to
temperature occur much faster in comparison to experiment. Towards the
freezing point, the density increases; however the rate of change does
not slow past 273.0 K. Evidently, \mdfts and \mccs do not feature the
expected density maximum around 277 K. This was also observed for the
SWM4-NDP and SWM3-HLJ models\cite{mackerell:2024} but not for q-AQUA
or MB-pol. Reasons for this behaviour include the neglect of 3- and
possibly 4-body interactions in \mdfts and \mcc. On the other hand, it
has been found that reproducing the density maximum with advanced
electrostatic models can be challenging and may require compromises in
performance on other desirable properties.\cite{mackerell:2024}
Finally, it is noted the only one out of 300 possible models from the
LJ-fit was assessed here and including $\rho(T)$ in the loss function
for optimizing the LJ-parameters may improve the performance.\\

\noindent
{\bf Thermodynamic Observables:} The enthalpy of vaporization ($\Delta
H_{\mathrm{vap.}}$) predicted by models \mdfts and \mccs are $10.3 \pm
0.1$ kcal/mol and $10.6 \pm 0.1$ kcal/mol, respectively, compared to
the experimental value of 10.51 kcal/mol (see
Table~\ref{tab:prop-train}). Simulations using the TIP3P, TIP4P, and
TIP5P models with constrained bonds and angles, yield values of 11.04,
11.15, and 10.73 kcal/mol, respectively.\cite{MM.dcm:2020} These
values tend to overestimate the experimental result as it is usually
preferred to better reproduce the temperature dependence of the
vaporization enthalpy.\cite{jenson:1998} For simulations run with
constraints such as SHAKE, the gas phase contribution to $\Delta
H_{\mathrm{vap.}}$ becomes constant, as such simulations neglect
vibrational contributions to thermodynamic properties. The
$T-$dependence of $\Delta H_{\rm vap}(T)$ as shown in Figure
\ref{fig:tempDep}F indicates that the trend in the cohesive energy per
molecule for higher temperatures is qualitatively correct but too weak
compared with experiment (solid line). Interestingly, a similar
intercept around 300 K in, and a negative slope of, the $\Delta H_{\rm
  vap}(T)$ curve was obtained using the AMOEBA-03 polarizable force
field.\cite{piquemal:2025} This could be a result of combination of
factors; critically neglecting many-body interactions in conjunction
with multipolar electrostatics, a design choice also made by kMDCM-NN.
In the case of water, these stronger electrostatics may contribute to
over-stabilizing the liquid in an ordered, low temperature state;
while over-estimating destabilizing interactions at higher
temperature, disordered states - particularly if many-body effects are
not well captured by the model. This points towards possible future
improvements in the nonbonded PES, specifically to the van der Waals
parameters.\\

\begin{table}[!t]
\centering
\caption{Test reference data: Thermodynamic, structural, and
  dielectric properties of water.  For kMDCM-NN, the $D_{0}$ was
  calculated using the $y$-intercept of the $D_{\mathrm{MD}}$
  vs. inverse side length plot.  $^{*}$Self diffusion coefficients for
  TIP3P, TIP4P, and TIP5P, were corrected with simulated shear
  viscosities taken from Mao \textit{et al.}\cite{Mao:2012} For
  $\tau_2$, the measured values were obtained from
  NMR\cite{Lankhorst:1982} or IR\cite{Rezus:2005} experiments,
  respectively.}  \scalebox{0.7}{
\begin{tabular}{clccccccc}
\toprule
&  & $\epsilon$ & $\Delta G$ (kcal/mol) & $\kappa$ ($10^6$ atm$^{-1}$) & $D_{\mathrm{MD}}$ ($10^{-5}$ cm$^2$/s) & $D_0$ ($10^{-5}$ cm$^2$/s) & $\tau_2$ (ps) \\
\midrule
 & Experiment & ~78.2 & $-6.3$ & ~45.8 $\pm$ 1.0 & - & ~2.36 $\pm$ 0.04 & $(1.7/2.5) \pm 0.1$ \\ \midrule
\multirow{2}{*}{kMDCM} & \mdft & ~63 & $-6.0 \pm 1.0$ & ~35.2 $\pm$ 1.5 & ~2.6 & ~2.9 $\pm$ 0.1 & $1.7 \pm 0.1$ \\
& \mcc & ~72 & $-6.2$ $\pm$ 0.1 & ~38.5 $\pm$ 1.1 & ~2.4 & ~2.6 $\pm$ 0.1 & 1.9 $\pm$ 0.1 \\
\midrule
\multirow{3}{*}{TIP} & TIP3P\cite{MM.dcm:2020} & ~103 & $-6.0 \pm 0.2$ & ~23.1 & ~3.9 & ~5.1$^{*}$ & ~1.1 \\
& TIP4P\cite{MM.dcm:2020} & ~59 & $-6.1 \pm 0.3$ & ~22.7 & ~2.3 & ~3.1$^{*}$ & ~1.8 \\
& TIP5P\cite{MM.dcm:2020} & ~91 & $-5.7 \pm 0.1$ & ~28.8 & ~2.2 & ~2.8$^{*}$ & ~2.3 \\
\midrule
Polarizable & AMOEBA\cite{Wang:2011, piquemal:2025} & 79 & -5.8  & ~46.6 $\pm$ 1.0 & ~2.0 & ~2.4 & ~2.2 \\
\bottomrule
\end{tabular}
}
\label{tab:prop-test}
\end{table}

\noindent
Table \ref{tab:prop-test} reports numerical values for additional
observables. The isothermal compressibility ($\kappa$) from
simulations using \mdfts and \mccs are $35.2 \pm 1.5 \times 10^{-6}$
and $38.5 \pm 1.1 \times 10^{-6}$ atm$^{-1}$, which underestimates the
experimental value of $45.8 \pm 1.0 \times 10^{-6}$ atm$^{-1}$. The
value obtained from iAMOEBA ($46.6 \pm 1.0 \times 10^{-6}$) is closer
to the literature value, although it was included as an objective in
the fitting procedure.\cite{Wang:2011} The agreement with experiment
is worse for the TIP models.\\

\noindent
The measured self-hydration free energy of water is $\Delta G_{\rm
  hyd} = -6.3$ kcal/mol based on the coexistence densities at 298
K.\cite{wagner:2002,weber:2010} This compares with computed values of
$-6.0 \pm 1.0$ and $-6.2 \pm 1.0$ kcal/mol for \mdfts and \mcc,
respectively, based on thermodynamic integration (Figure
\ref{sifig:dGwater}). As expected for such a polar solvent, the
electrostatic terms contribute most to the stabilizing value of
$\Delta G$ of self-hydration. Literature values for the TIP family of
models compared in Table \ref{tab:prop-test} report similar
agreements, all roughly within the predicted uncertainty. On the other
hand, the i-AMOEBA model underestimates the self-hydration free energy
of water by roughly 1 kcal/mol which does not change even when
switching to a fluctuating charge model.\cite{piquemal:2025}\\

\noindent
{\bf Transport Properties:} Orientational lifetimes of the O-H bond
vectors, as well as self-diffusivities, were calculated from longer 20
ns (two repeats) $NVE$ trajectories using both parameter sets and
compared to literature values for TIP3P, TIP4P, and TIP5P (Table
\ref{tab:prop-test}). Simulations with 2000 and 8000 water molecules
were used to include finite size corrections to the self-diffusion
coefficient $D_{\mathrm{MD}}$ calculated with periodic boundary
conditions, to compare with the experimental value
$D_{\infty}$. Corrections to the reported $D_{\mathrm{MD}}$ for the
TIP family of water models were added using the calculated viscosities
from the literature.\cite{Mao:2012} The self-diffusion coefficient for
\mdfts ($2.9 \pm 0.1 \cdot 10^{-5}$ cm$^2$/s) is closer to the
corrected values obtained using TIP5P ($2.8 \cdot 10^{-5}$ cm$^2$/s)
and \mccs is closest to the experimental value of $2.35 \pm 0.04 \cdot
10^{-5}$ cm$^2$/s measured with pulsed gradient NMR.\cite{Holz:2000}
The correction described in Eq. \ref{eq:Dcorr} implies a shear
viscosity, $\eta$, of $\sim 2.18 \cdot 10^{-4}$ kg m$^{-1}$ s$^{-1}$,
roughly 75\% of the experimental value. Adjusting the apparent
self-diffusion coefficient by this ratio results in a value of $2.2
\cdot 10^{-5}$ cm$^2$/s. This suggests that the transport behaviour of
the model is qualitatively consistent with the experiment when
accounting for these factors. For TIP3, the corrected self-diffusion
coefficient of $5.1 \cdot 10^{-5}$ cm$^2$/s is much too high.\\

\begin{figure}[h!]
    \centering
    \includegraphics[width=.5\linewidth]{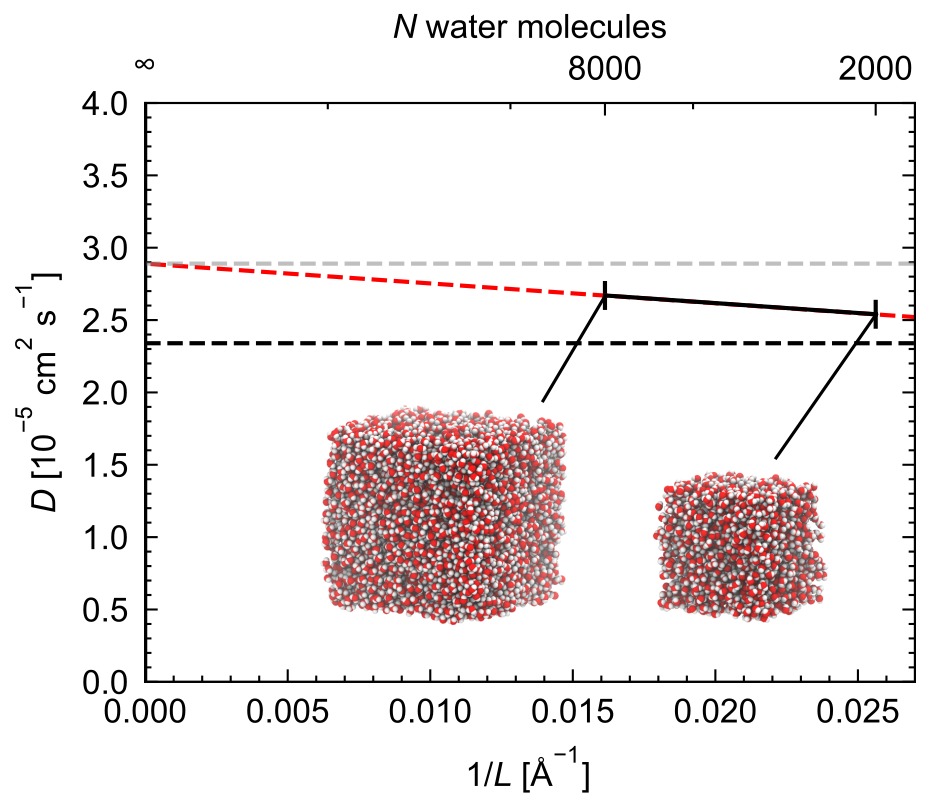}
    \caption{To account for finite-size effects, the self-diffusion
      coefficient obtained from length scales accessed by simulation,
      $D_{\mathrm{PBC}}$, must be extrapolated (grey) in the limit of
      infinite side length, $L$, in order to be compared to the
      experimental value, $D_{\infty}$ ( black horizontal line).}
    \label{fig:Dinf}
\end{figure}

\noindent
As a last observable, the rotational life time $\tau_2$ were
determined as $\tau_2^{\rm \mdft} = 1.7 \pm 0.1$ ps and $\tau_2^{\rm
  \mcc} = 1.9 \pm 0.1$ ps. This compares with measured values of 1.7
ps and $2.5 \pm 0.1$ ps from proton NMR relaxation of H$^{17}_2$O and
from pump-probe spectroscopy,
respectively.\cite{Lankhorst:1982,Rezus:2005} Hence, both models are
closer to the results from NMR experiments but still consistent with
findings from IR spectroscopy.\\

\begin{figure}[h!]
\centering
\includegraphics[width=0.7\textwidth]{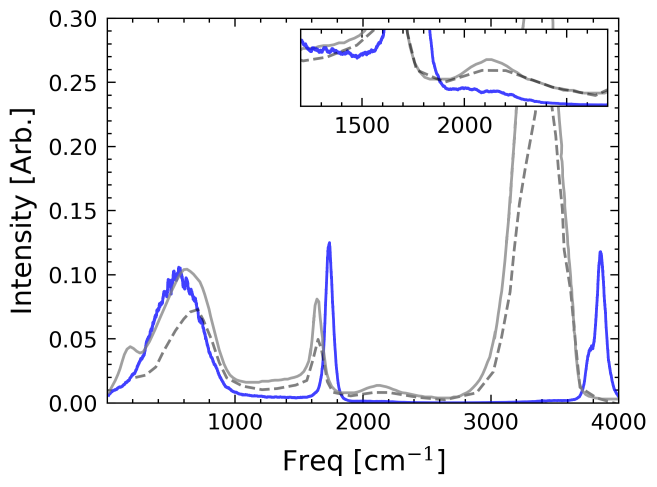}
\caption{Infrared spectrum of liquid water between 0 and 4000
  cm$^{-1}$ obtained from 1 ns simulation using the \mdfts parameters
  at 300~K in the $NVT$ ensemble (blue trace). The experimentally
  reported spectra are the grey solid\cite{bertie:1996} and
  dashed\cite{chapados:2009} lines.}
\label{fig:ffnet_ir}
\end{figure}

\noindent
{\bf Vibrational Spectroscopy} Finally, the calculated and measured
condensed-phase IR spectra for water are shown in
Figure~\ref{fig:ffnet_ir}. While the main features of the experiment
are reproduced by the simulations, the bending and OH stretch peaks
remain blue shifted as expected. The general shapes, the broadness and
the intensities are in good agreement in particular for the two peaks
below 2000~cm$^{-1}$. There are two measured signals below $\sim 200$
and 2150~cm$^{-1}$, which are not captured in the MD simulations. The
former, a hydrogen-bond stretching peak, is entirely absent in the
present calculation and is likely due neglect of intermolecular charge
transfer.\cite{sidler:2018,yang2024chargetransfer,han2023incorporating}
The band at 2150~cm$^{-1}$ corresponds to a bending-libration
combination mode, which is also underestimated in intensity. Comparing
the experimental gas- and liquid-phase peak positions of the bending
mode (see Table~\ref{tab:harm_freq_water_ccf12}) reveals a blue-shift
of $\sim 55$~cm$^{-1}$. This blue-shift is overestimated in the
calculations ($105$~cm$^{-1}$). In contrast, incorporating
anharmonicity via VPT2 yields frequencies that quantitatively agree
with experimental values, which emphasizes the quality of the bonded
PES.\\

\section{Discussion and Outlook}
This manuscript introduces a generic ML-based parametrization strategy
for condensed phase systems. The internal monomer contributions are
represented as a kernel-based NN, for the anisotropic and fluctuating
electrostatics a kernel-based distributed charge model is used, and
Lennard-Jones contributions were fitted to pure cluster data. Among
the multiple solutions of the optimization problem for the van der
Waals interactions those providing closest agreement with experimental
density and heat of vaporization were chosen. This protocol was
applied to water as a case study for which ample experimental and
computational results are available for validation. It is demonstrated
that this approach yields qualitatively correct to very good
thermodynamic properties compared with experiment not included in the
parametrization. The models allow efficient and energy conserving
simulations and were used for routine 10 ns MD simulations for water
box sizes up to $\sim 10^4$ monomers using commodity hardware.\\

\noindent
Experimentally measured pure liquid densities and heats of
vaporization are available for a wide range of
systems.\cite{crc:2023,refprop:2022} Similarly, measured gas-phase
vibrational spectra for validating the monomer energy functions are
also typically known. Within the realm of the present approach this
information was found to provide practical constraints for the
LJ-parameters. Such an empirical force field-inspired approach for a
machine learning-based parametrization thus consists of the following
elements:
\begin{itemize}
\item Accurate monomer energy functions can be determined efficiently
  from a wide range of machine learning-based techniques, including
  kernel- or neural network-based approaches.\cite{yu:2022,zhu2023mb}
  Achievable levels of theory are at least Moller-Plesset second order
  theory (MP2) but capitalizing on transfer-learning techniques,
  higher levels of theory including CCSD(T) levels are within reach as
  well.\cite{MM.tl:2022}
\item For the electrostatic contributions a wide array of techniques
  are available, ranging from multipolar to distributed charge
  techniques.
 \item Finally, LJ-parameters are determined from electronic structure
   calculations for finite-size cluster interaction energies. For
   sufficiently large clusters ($N \sim 10$ monomers) the only
   available techniques at present are based on density functional
   theory, though.
\end{itemize}
In a next step, which was not considered here, the LJ-parameters can
be adjusted to better capture additional condensed-phase properties
such as $g(r)$, $\kappa$, $D$ and others together with their
$T-$dependence depending on available measurements. In this fashion a
closed-cycle loop combining high(est)-level electronic structure
reference data, machine learning techniques for best representing the
electronic structure data, cluster models for two- and many-body
interactions between monomers, and measured information for model
selection emerges that can be applied to systems for which the
necessary experimental data is available.\\

\noindent
For the specific case of ``water'' considered in the present work, the
two-body contributions perform very satisfactorily when compared with
two-body CCSD(T)/CBS data. However, higher order many-body
interactions were neglected and including them will further boost
model performance. The contributions of three- and four-body terms
have been estimated to be less than 2\% and 0.5\% of the total
interaction energy, respectively.\cite{reddy:2016}. An interesting
detail concerns the dependence of the model on the level of theory of
the underlying QM calculation, see
Figure~\ref{sifig:clusterfitsim}. Models fitted to interaction
energies of increasingly large clusters in the reference data set
perform best on $[\rho, \Delta H_{\rm vap}]$ for cluster sizes $N \in
[2,10]$. This finding illustrates the along a parametrization effort
various checks can and should be carried out to ascertain model
performance. \\

\noindent
As a road-map for further improvements of the \mdfts and \mccs models
described here, it is valuable to compare the TIP4P and TIP4P/2005
models. The original TIP4P model\cite{jorgensen:1983} has partial
charges on the hydrogen atoms and a negative charge on a dummy (M)
site along the bisector of the H–O–H angle, while the LJ
interaction site is centered on the oxygen atom. Although TIP4P
reproduces the density of liquid water near ambient conditions
reasonably well, the model shows significant deviations from
experiment in properties such as $\rho(T)$, the melting point of ice,
diffusion coefficients, and surface tension. To improve upon these
deficiencies, the TIP4P/2005 model\cite{abascal:2005} was introduced
which preserves the four-site geometry of TIP4P but re-optimizes key
parameters to better reproduce a wide range of experimental properties
across multiple phases of water. First, the negative charge on the
M-site was increased from $-1.0400 e$ in TIP4P to $-1.1128 e$ in
TIP4P/2005 with concomitant changes in the H-atom charges ($+0.52 e$
to $+0.5564 e$). Next, the M-site, originally positioned 0.15~\AA\/
from the oxygen atom along the H--O--H bisector in TIP4P, was adjusted
to 0.1546~\AA\/ in TIP4P/2005. Finally, the depth and size of the
LJ-potential change from [0.1550 kcal/mol / 3.15365 \AA\/] to [0.1852
  kcal/mol / 3.1589 \AA\/].\\

\noindent
These adjustments yield significantly improved agreement with
experiment across a wide range of properties. For example, TIP4P/2005
correctly predicts the experimental density maximum near 277~K, and
increases the predicted melting point of ice to approximately 250~K
(compared to $\sim 230$~K for TIP4P). The self-diffusion coefficient
of liquid water at 298~K is $3.5 \times 10^{-5}$~cm$^2$/s ($2.3 \times
10^{-5}$~cm$^2$/s with finite-size corrections) for TIP4P/2005
(experiment: $2.3 \times 10^{-5}$~cm$^2$/s) which is a significant
improvement over $5.9 \times 10^{-5}$~cm$^2$/s from TIP4P. Finally,
the surface tension at 298~K is 70.2~mN/m for TIP4P/2005 (experiment:
71.7~mN/m) compared to 61~mN/m for TIP4P. On the other hand, the
enthalpy of vaporization at 298~K is 10.6~kcal/mol for TIP4P, in good
agreements with the experimental value of 10.52~kcal/mol, compared
with 11.99~kcal/mol for TIP4P/2005. Along similar lines it is
anticipated that including additional experimental constraints in the
parametrization will improve the \mdfts and \mccs models.\\

\noindent
The present work underscores that further progress is needed in
modeling van der Waals interactions. Ideally, a separate ML-based
representation will be used for this. Another future improvement is to
go beyond isotropic van der Waals models, akin to anisotropic
electrostatic interactions. Finally, the challenge will be to balance
performance and accuracy of such models so that they can still be used
for stable and accurate large-scale simulations for condensed-phase
systems.\\

\noindent
In summary, as was demonstrated here, high-level electronic structure
calculations, machine learning-based representations, and refinement
based on comparison between experimental and computed condensed-phase
properties is a robust and cost-effective approach for rational and
purpose-targeted construction of energy functions. Deliberately
restraining LJ-parameter selection on performance for $[\rho, \Delta
  H_{\rm vap}]$, which are available for a wide array of liquids,
\mdfts and \mccs models were found to perform satisfactorily on a wide
range of experimental observables. Including some of these properties
together with their $T-$dependence is expected to further improve
model performance within the limit of the LJ-representation for van
der Waals interactions. For water, routine extended-system simulations
at levels near or equivalent to CCSD(T)-quality are possible. Further
improvements concern inclusion of 3- and higher many-body interactions
(which are not relevant for all liquids) and new parametrization
strategies for van der Waals interactions. It is anticipated that such
coherent parametrization workflows contribute in the future to improve
also the parametrization and simulation of heterogeneous and
electrostatically interesting systems, such as (deep) eutectic or
ionic liquids.\\

\section*{Data Availability}
The data accompanying this work is available at
\url{https://github.com/MMunibas/water.kernn.mdcm}. Also, the python
code for the Schlegel hypercube analysis is available at this
repository.

\section*{Acknowledgment}
We thank the Swiss National Science Foundation (grants
200020{\_}219779 and 200021{\_}215088) (to MM), and the University of
Basel for supporting this work. The authors thank the MacKerell group
for providing trajectories from SWM4-HLJ simulations.\\

\bibliography{references}

\clearpage

\renewcommand{\thetable}{S\arabic{table}}
\renewcommand{\thefigure}{S\arabic{figure}}
\renewcommand{\thesection}{S\arabic{section}}
\renewcommand{\d}{\text{d}}
\setcounter{figure}{0}  
\setcounter{section}{0}  
\setcounter{table}{0}

\newpage

\noindent
\LARGE{\bf SUPPORTING INFORMATION:\\ Towards Large-Scale Condensed Phase
  Simulations using Machine Learned Energy Functions}

\newpage

\section{Figures}

\begin{figure}[h!]
\centering
\includegraphics[width=1.0\textwidth]{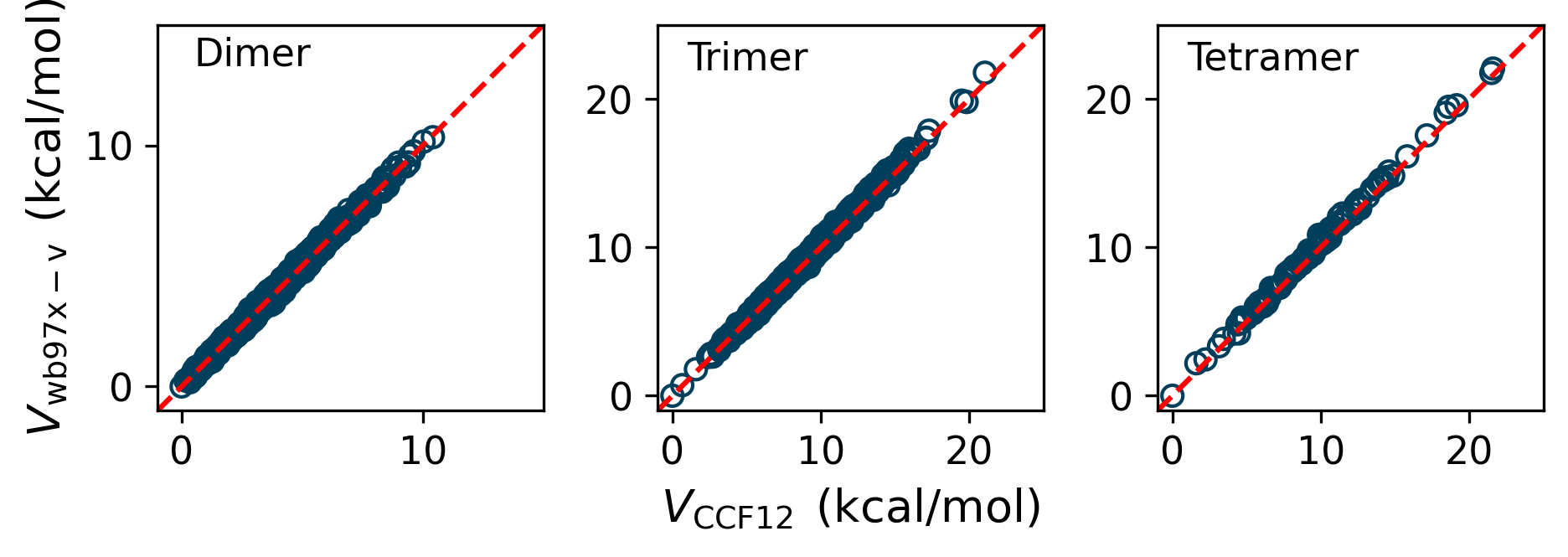}
\caption{Comparison of \textit{ab initio} CCSD(T)-F12/aug-cc-pVTZ and
  $\omega$B97X-V/def2-QZVP energies for (H$_2$O)$_2$ to
  (H$_2$O)$_4$. The energies are normalized with respect to the
  cluster with lowest energy. For dimers and trimers, a total of 1000
  configurations are compared while for tetramers only 100 structures
  served as a benchmark.}
\label{sifig:dft_vs_ccf12_benchmark}
\end{figure}

\begin{figure}[h!]
    \centering
    \includegraphics[width=0.9\linewidth]{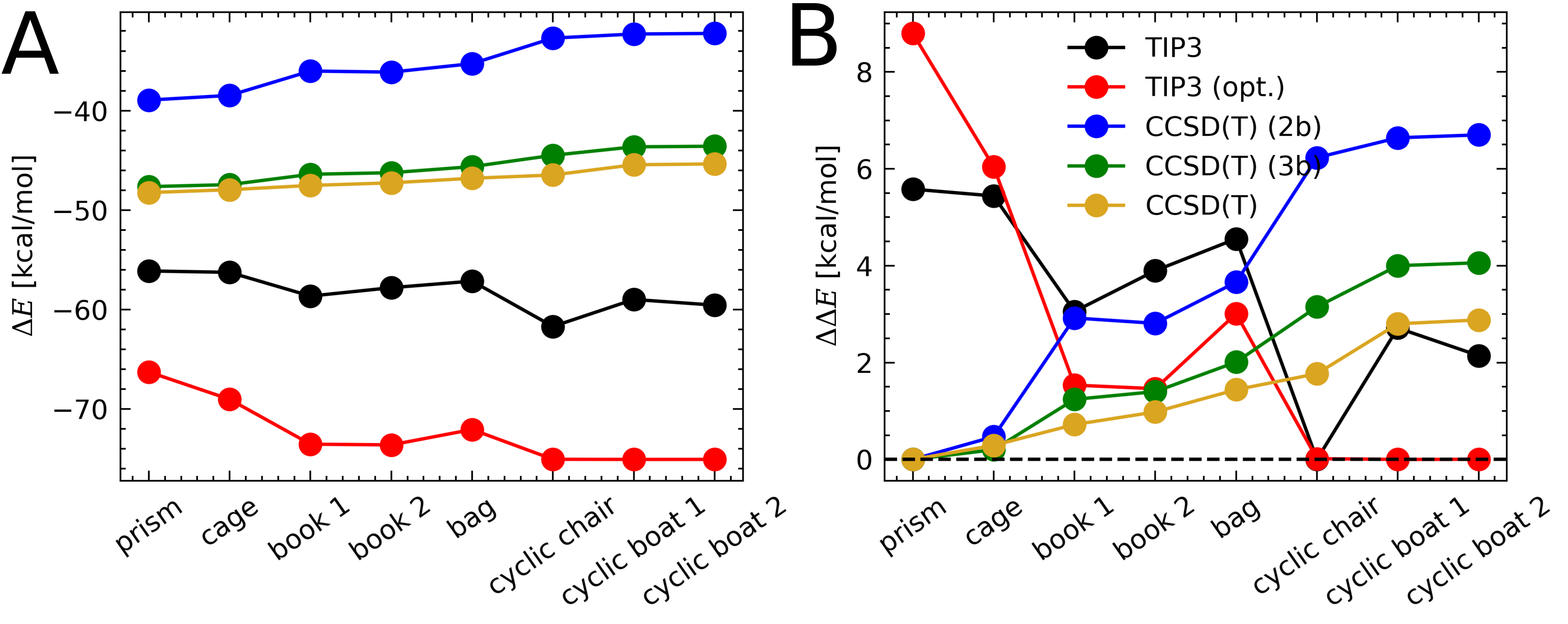}
    \caption{(A) Interaction energies of all low-lying isomers of
      the (H$_2$O)$_6$ clusters as obtained through optimization with
      the CHARMM TIP3 model. CCSD(T) data taken from
      Ref.\citenum{reddy:2016}, including contributions up to two-body
      (2b) and three-body (3b) interactions. (B) Interaction energies
      relative to the lowest energy conformer.}
    \label{sifig:hexamers_tip3}
\end{figure}

\begin{figure}[h!]
    \centering
    \includegraphics[width=1.0\linewidth]{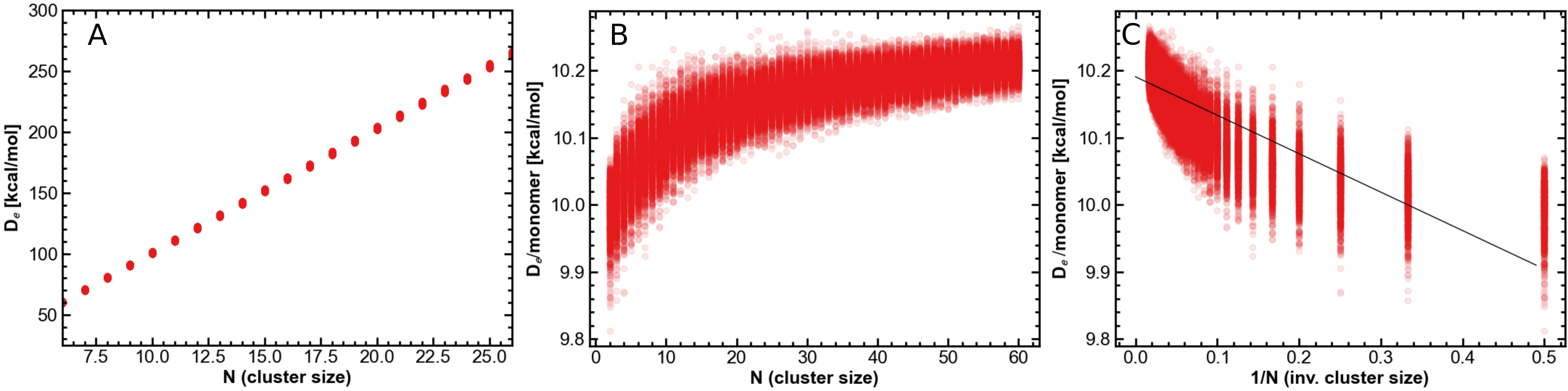}
    \caption{(A) Electronic dissociation energies ($D_e$) per cluster
      size and (B) per monomer, and (C) as a function of inverse
      cluster size which predicts $D_e$/monomer of 10.2 kcal/mol in
      the limit of increasing cluster sizes.}
    \label{sifig:sizeDe}
\end{figure}

\begin{figure}[h!]
    \centering
    \includegraphics[width=0.5\linewidth]{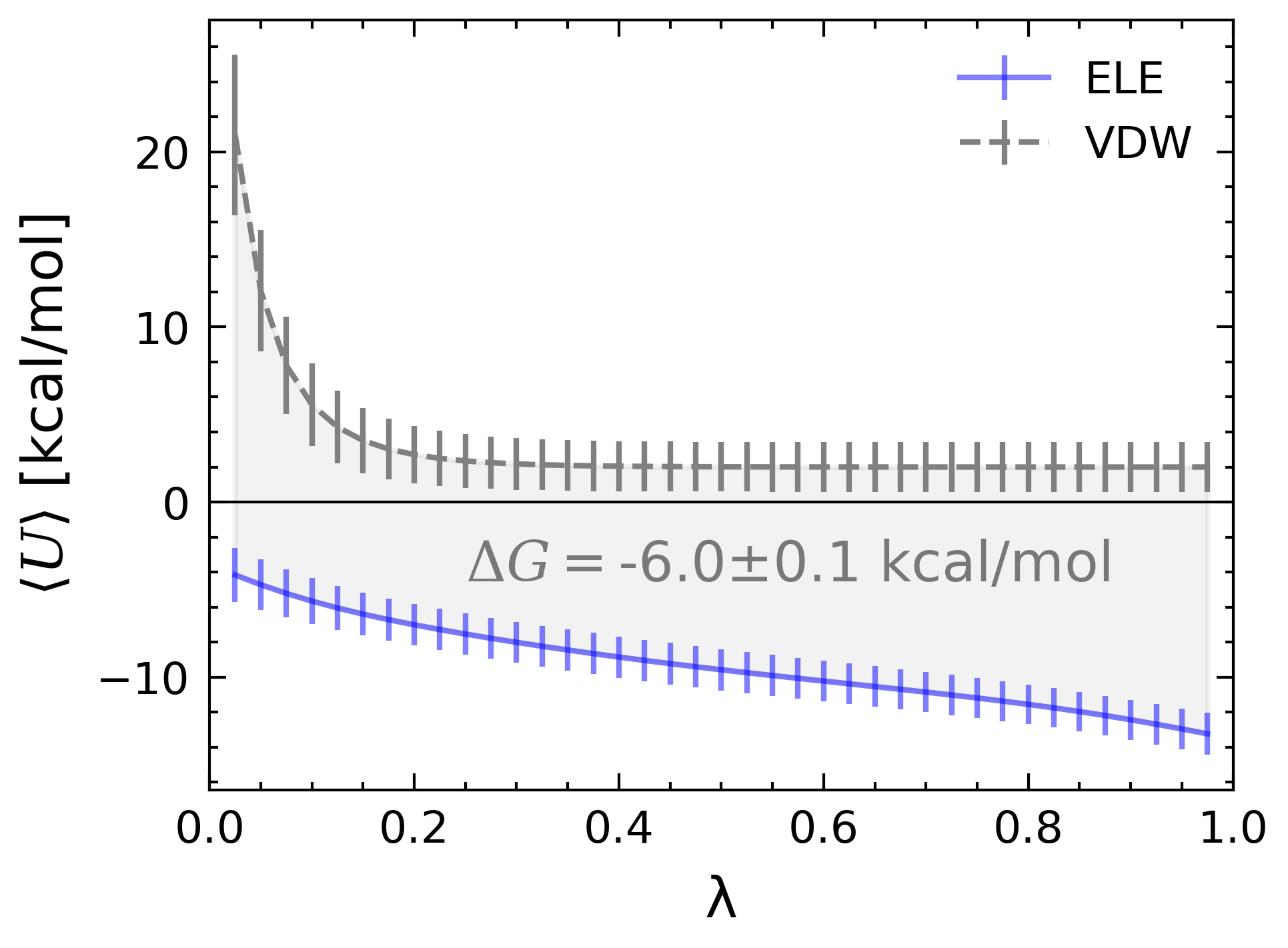}
    \caption{\textbf{M-DFT:} The free energy of hydration calculated
      using thermodynamic integration over 36 $\lambda$ windows. }
    \label{sifig:dGwater}
\end{figure}

\begin{figure}[h!]
    \centering
    \includegraphics[width=0.75\linewidth]{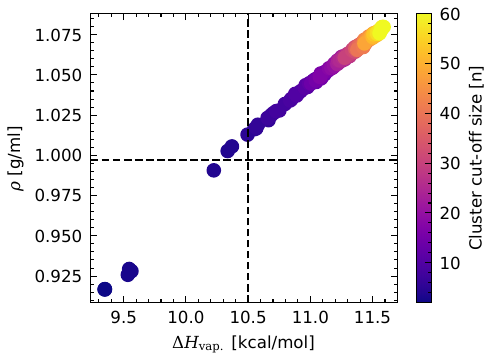}
    \caption{\mdft: Density ($\rho$) and heat of vaporization $\Delta
      H_{\mathrm{vap.}}$ from $CpT$ simulations using refined LJ
      parameters based on fitting to water clusters of varying
      sizes. For each size cut-off, three parameter sets from
      optimizations using different random seeds were used.}
    \label{sifig:clusterfitsim}
\end{figure}
\clearpage

\section{Parameters}
 
\begin{table}[h!]
    \centering
    \renewcommand{\arraystretch}{1.2}
    \begin{tabular}{lcccc}
        \toprule
        \textbf{Parameter} & \textbf{OT (M-DFT)} & \textbf{HT (M-DFT)} & \textbf{OT (M-CC)} & \textbf{HT (M-CC)} \\
        \midrule
        $\boldsymbol{\epsilon}$ (kcal/mol) & -0.2275 & -0.2470 & -0.1125 & -0.0434 \\
        $\sigma$ (\AA) & 1.7653 & 0.0230 & 1.8194 & 0.4265 \\
        \bottomrule
    \end{tabular}
    \caption{Lennard-Jones parameters for final models \mdft~ and \mcc.}
    \label{sitab:atom_params}
\end{table}

\end{document}